\documentclass[preprint]{aastex631}

\usepackage{soul}

\shorttitle{HAT-P-37b: TTV and Atmosphere}
\shortauthors{N.A-thano et al.}
\graphicspath{{./}{figures/}}
\begin{document}

\title{The Transit Timing and Atmosphere of Hot Jupiter HAT-P-37b}

\author[0000-0001-7234-7167]{Napaporn A-thano}
\affiliation{Department of Physics and Institute of Astronomy, National Tsing-Hua University, Hsinchu 30013, Taiwan}
\email{napaporn@gapp.nthu.edu.tw}

\author[0000-0001-7359-3300]{Ing-Guey Jiang}
\affiliation{Department of Physics and Institute of Astronomy, National Tsing-Hua University, Hsinchu 30013, Taiwan}
\email{jiang@phys.nthu.edu.tw}

\author[0000-0003-3251-3583]{Supachai Awiphan}
\affiliation{National Astronomical Research Institute of Thailand, 260 Moo 4, Donkaew, Mae Rim, Chiang Mai, 50180, Thailand}
\email{supachai@narit.or.th}

\author{Ronnakrit Rattanamala}
\affiliation{PhD Program in Astronomy, Department of Physics and Materials Science,\\ Faculty of Science, Chiang Mai University, Chiang Mai, 50200, Thailand}
\affiliation{Department of Physics and General Science, Faculty of Science and Technology, \\ Nakhon Ratchasima Rajabhat University, Nakhon Ratchasima , 30000, Thailand}

\author{Li-Hsin Su}
\affiliation{Department of Physics and Institute of Astronomy, National Tsing-Hua University, Hsinchu 30013, Taiwan}

\author{Torik Hengpiya}
\affiliation{Regional Observatory for the Public, Songkhla, 79/4 Moo 4, Khao Roop Chang, Muang District, Songkhla, 90000, Thailand}

\author{Devesh P. Sariya}
\affiliation{Department of Physics and Institute of Astronomy, National Tsing-Hua University, Hsinchu 30013, Taiwan}

\author{Li-Chin Yeh}
\affiliation{Institute of Computational and Modeling Science, National Tsing-Hua University, Hsinchu 30013, Taiwan}

\author{A. A. Shlyapnikov}
\affiliation{Crimean Astrophysical Observatory, 298409, Nauchny, Crimea}

\author{Mark A. Gorbachev}
\affiliation{Crimean Astrophysical Observatory, 298409, Nauchny, Crimea}

\author{Alexey N. Rublevski}
\affiliation{Crimean Astrophysical Observatory, 298409, Nauchny, Crimea}

\author{Vineet Kumar Mannaday}
\affiliation{
Department of Pure \& Applied Physics,
Guru Ghasidas Vishwavidyalaya (A
Central University), Bilaspur (C.G.)-495009, India}

\author{Parijat Thakur}
\affiliation{
Department of Pure \& Applied Physics,
Guru Ghasidas Vishwavidyalaya (A
Central University), Bilaspur (C.G.)-495009, India}

\author{D. K. Sahu}
\affiliation{Indian Institute of Astrophysics, Bangalore–560034, India}

\author{David Mkrtichian} 
\affiliation{National Astronomical Research Institute of Thailand, 260 Moo 4, Donkaew, Mae Rim, Chiang Mai, 50180, Thailand}

\author{Evgeny Griv} 
\affiliation{Department of Physics, Ben-Gurion University, Beer-Sheva 84105, Israel}



\begin{abstract}
The transit timing variation (TTV) and transmission spectroscopy analyses of the planet HAT-P-37b,
which is a hot Jupiter orbiting an G-type star, were performed.
Nine new transit light curves are obtained and analysed together with 21 published light curves from the literature. The updated physical parameters of HAT-P-37b are presented. The TTV analyses show a possibility that the system has an additional planet which induced the TTVs amplitude signal of 1.74 $\pm$ 0.17 minutes. If the body is located near the 1:2 mean motion resonance orbit, the sinusoidal TTV signal could be caused by the gravitational interaction of a sub-Earth mass planet with mass of 0.06 $M_\oplus$. From the analysis of an upper mass limit for the second planet, the Saturn mass planet with orbital period less than 6 days is excluded. The broad-band transmission spectra of HAT-P-37b favours a cloudy atmospheric model with an outlier spectrum in $B$-filter. 
\end{abstract}



\section{Introduction} \label{sec:intro}
Discoveries of new extra-solar planets through the transit method has been dramatically grown in recent years. More than 3,000 planets\footnote{The Extrasolar
Planets Encyclopaedia: \texttt{http://exoplanet.eu/}} have been confirmed by transit method. Since the launch of \textit{Kepler} space telescope in 2009, more than 2,600 planets have discovered using the \textit{Kepler} data \citep{borucki2010}. After the \textit{Kepler} era, the grand of exoplanet detection by transit method has been developed by the Transiting Exoplanet Survey Satellite (\textit{TESS}; \citet{ricker2014}). To date, more than 140 planets have confirmed by the \textit{TESS} mission. Transit light curves can be used to search for additional planets in planetary system via the transit timing variations (TTVs) \citep{agol2005,holman2005,macie2010,jiang2013}. Additionally, the TTV signal can be used to examine the theoretical predictions of orbital period changes, orbital decay and apsidal precession (see \citet{Maciejewski2016,patra2017,southworth2019,mannaday2020}).

In addition to the discovery of new exoplanets and investigating planetary dynamics, the characterization of planetary interiors and atmospheres is a rapidly developing area. One method that is used to study planetary atmospheres is transmission spectroscopy, which measures the variation of transit depth with wavelength \citet{seager2000}. The technique has been proven to be one of the most powerful techniques to characterize planet atmospheres. The first planet atmospheres modeling provided by \citet{seager2000}. The first high precision spectro-photometric observations of HD 209458 through the absorption from sodium in the planetary atmosphere with \textit{Hubble Space Telescope} (\textit{HST}) was reported by \citet{charbon2002}. \citet{sing2016} demonstrated the comparative studies of ten hot Jupiters' atmospheres using transmission spectroscopy. They found that the difference between the planetary radius measured at optical and infrared wavelengths can be applied for distinguishing different atmosphere types.  

The Hungarian-made Automated Telescope Network (HATNet) is a ground-based telescope network of seven wide-field small telescopes, which monitor bright stars from $r\approx 9$ mag to $r\approx 14$ mag in order to search for new exoplanets via transit method \citet{bakos2004}\footnote{The Hungarian-made Automated Telescope Network: \texttt{https://hatnet.org/}}. Since the first light in 2003, the HATNet survey has discovered 70 extra-solar planets. A number of hot Jupiters, including HAT-P-58b - HAT-P-60b \citep{bakos2021} and HAT-P-35b - HAT-P-37b \citep{bakos2012}, were discovered by the surveys. In this work, we focus on the photometric follow-up observations of a hot Jupiter, HAT-P-37b.   

HAT-P-37b, a hot Jupiter orbiting around the host G-type star HAT-P-37 ($V$ = 13.2, $M_\star$ = 0.929 $\pm$ 0.043$M_\odot$, $R_\star$ = 0.877 $\pm$ 0.059$R_\odot$, $T_{eff\star}$ = 5500 $\pm$ 100 K and $log_g\star$ = 4.67 $\pm$ 0.1 cgs) with a period of 2.8 days, was discovered by \citet{bakos2012}. The existence of HAT-P-37b has been confirmed by radial velocity measurements from high-resolution spectroscopy using the Tillinghast Reflector Echelle Spectrograph (TRES) and three $i'$ band follow-up light curves from KeplerCam instrument. From the data, HAT-P-37b is a Jupiter-mass exoplanet with mass $M_p$ = 1.169 $\pm$ 0.103$M_{Jup}$, radius $R_p$ = 1.178 $\pm$ 0.077$R_{Jup}$ and equilibrium temperature $T_{eq}$ = 1271 $\pm$ 47 K. 

In 2016, HAT-P-37b was revisited by \citet{Maciejewski2016}. The obtained planetary parameters modelled from four new transit light curves and published light curves from \citet{bakos2012} are consistent with the values in \citet{bakos2012} paper. \citet{turner2017} presented the photometric follow-up observation by the 1.5-m Kuiper Telescope with $R$ and $B$ band. They derived the physical parameters by combination of their two transit light curves and previous public data. They found that the transit depth in $B$ band is smaller than the depth in near-IR bands. The variation may be caused by the TiO/VO absorption in the HAT-P-37b's atmosphere. Recently in 2021, \citet{yang2021} performed follow-up photometric observations of HAT-P-37b using the 1-m telescope at Weihai Observatory. The physical and orbital parameters of HAT-P-37b are refined by their new nine light curves combined with published data. The investigation of dynamic analysis was presented and there was no significant of TTV signal from new ephemeris given rms scatter of 57 second.  

In this work, we present new ground-base photometric follow-up observations of 9 transit events of HAT-P-37b. These data are combined with available published photometric data in order to constrain the planetary physical parameters, investigate the planetary TTV signal, and constrain the atmospheric model. Our observational data are presented in Section \ref{sec:observation}. The light curves analysis is described in Section \ref{sec:LCanalysis}. The study of TTVs includes timing models, the frequency analysis, and the upper mass limit of additional planets, as presented in Section \ref{sec:ttv}. In Section \ref{sec:atmosphere}, the analysis of HAT-P-37b atmosphere is given. Finally, the discussion and conclusion are in Section \ref{sec:conclusion}.

\section{Observational Data}
\label{sec:observation}
\subsection{Observations and data reduction}
\label{subsec:observe}
The photometric observations of HAT-P-37b were conducted using 60-inch telescope (P60) at Palomar Observatory, USA, the 50-cm Maksutov telescope (MTM-500) at the Crimean Astrophysical Observatory, Crimea, and 0.7-m Thai Robotic Telescope at Sierra Remote Observatories , USA, between 2014 June and 2021 July. Nine transits, including five full transits and four partial transits, in $R$-band and $B$-band are obtained. The observation log is given in Table~\ref{tab:1}

\emph{$\bullet$ The 60-inch telescope (P60)}  \\
\indent One full transit and three partial transits of HAT-P-37b were obtained by the 60-inch telescope (P60) at Palomar Observatory, California, USA in 2014. The P60 is a reflecting telescope built with Ritchey–Chr$\acute{e}$tien optics. The field of view of each image is 13 $\times$ 13 arcmin$^{2}$, with a 2048 $\times$ 2048 pixels CCD camera. 

\emph{$\bullet$ The 50-cm Maksutov telescope (MTM-500)} \\
\indent During 2017-2020, three full transits of HAT-P-37b are obtained with with the 50-cm Maksutov telescope (MTM-500) at the Crimean Astrophysical Observatory (CrAO), Nauchny, Crimea. The observations were perform using an Apogee Alta U6 1024 $\times$ 1024 pixels CCD camera. The field of view is about 12 $\times$ 12 arcmin$^{2}$.

\emph{$\bullet$ 0.7-m Thai Robotic Telescope at Sierra Remote Observatories (TRT-SRO)} \\
\indent Recently, One full transit and one partial transit are obtained by the 0.7-m telescope is a part of Thai Robotic Telescope Network operated by National Astronomical Research Institute of Thailand (NARIT). The 0.7-m Robotic Telescope is located at Sierra Remote Observatories (TRT-SRO), California, USA. We observed HAT-P-37b with the Andor iKon-M 934 1024 $\times$ 1024 pixels CCD camara. The field of view of 10 $\times$ 10 arcmin$^{2}$

\emph{$\bullet$ Data Reduction} \\
\indent All the science images of HAT-P-37b were calibrated by bias-subtraction, dark-subtraction and flat-corrections using the standard tasks from {\tt\string IRAF}\footnote{IRAF is distributed by the National Optical Astronomy Observatories, which are operated by the Association of Universities for Research in Astronomy, Inc., under cooperative agreement with the National Science Foundation. For more details, \texttt{http://iraf.noao.edu/}} package. The astrometic calibration for all science images were perform by {\tt\string Astrometry.net} \citep{lang2010}. To create the transit light curve for each observation, the aperture photometry was carried out using {\tt\string sextractor} \citep{berlin1996}.The nearby stars with magnitude $\pm$ 3 from HAT-P-37b without strong brightness variations were selected to be the reference stars. The time stamps are converted to Barycentric Julian Date in Barycentric Dynamical Time (BJD$_\textup{TDB}$) using {\tt\string barycorrpy} \citep{Kano2018}. 

\subsection{Literature data}
In order to obtain the HAT-P-37b parameters, 9 transit light curves in Section \ref{subsec:observe} are combined with 21 published transit light curves. These published light curves include 3 $i'$-band light curves from \citet{bakos2012}, 4 light curves from \citet{Maciejewski2016} (2 in Cousins $R$-band, 1 in Gunn-$r$ band and 1 with no filter), 2 light curves in Harris $B$ and $R$ filters from \citet{turner2017}, 3 $R$-band light curves from \citet{wang2021}, and 9 light curves which includes 7 in $V$-band, 2 in $R$-band from \citet{yang2021}. In total, 30 transit light curves of the HAT-P-37b are used in this work.

Note that we have checked HAT-P-37 data from the Kepler/K2 and TESS databases.
HAT-P-37 is not in the Kepler/K2 fields. The planetary system was observed
by TESS. However, there is a bright nearby binary system ($<$ 5 TESS pixels).
Therefore, the HAT-P-37 TESS light curves are diluted with the flux from
that nearby binary and we could not easily detect the HAT-P-37b transit.
Therefore, we didn't include TESS light curves in this work.

\begin{table*}
\begin{center}
\caption{Log of observations of HAT-P-37b transits. Epoch=0 is the transit on 2011 March 21.}
\label{tab:1}          
\small\addtolength{\tabcolsep}{-2pt}
\begin{tabular}{lccccccc}
\toprule
Observation Date  & Epoch & Telescope  & Filter & Exposure time (s) & Number of Images & PNR (\%)& Transit coverage \\
\hline
2014 May 28     & 416   & P60     & $R$  & 30    & 68   & 0.24   & Ingress only  \\ 
2014 June 11    & 421   & P60     & $R$  & 30    & 101  & 0.12   & Ingress only   \\
2014 July 23    & 436   & P60     & $R$  & 30    & 160   & 0.15  & Full          \\   
2014 August 06  & 441   & P60     & $R$  & 30    & 92    & 0.14   & Ingress only  \\  
2017 April 02   & 788   & MTM-500  & $R$  & 60    & 180  & 0.39   & Full     \\ 
2019 April 05   & 1050  & MTM-500  & $R$  & 60    & 147  & 0.40   & Full     \\ 
2020 July 18    & 1218  & MTM-500  & $R$  & 60    & 149  & 0.49   & Full     \\ 
2021 July 20    & 1349  & TRT-SRO  & $R$    & 30   & 405 & 0.37   & Full     \\  
2021 August 03  & 1354  & TRT-SRO  & $B$    & 90   & 95  & 1.25   & Egress only     \\ 
\hline
\end{tabular}
\end{center}
{Notes: PNR is the photometric noise rate \citep{Fulton2011}.}
\end{table*}

\begin{table*}
\begin{center}
\caption{The detrended photometric data of HAT-P-37b transits in this work. The 2nd order polynomial detrending functions in \texttt{TransitFit} were used. Epoch=0 is the transit on 2011 March 21.}
\label{tab:2}          
\small\addtolength{\tabcolsep}{-2pt}
\begin{tabular}{lcccc}
\toprule
Epoch & BJD & Normalized Flux & Normalised flux   \\
      &     &                 & uncertainty       \\
\hline
416	&	2456805.80023	&	0.999	&	0.005	\\
	&	2456805.80087	&	0.999	&	0.005	\\
	&	2456805.80151	&	1.000	&	0.005	\\
	&	2456805.80216	&	1.001	&	0.005	\\
	&	2456805.80344	&	1.002	&	0.005	\\
	&	…	&	…	&	…	\\
\hline							
421	&	2456819.81466	&	1.001	&	0.003	\\
	&	2456819.81531	&	0.999	&	0.003	\\
	&	2456819.81595	&	1.000	&	0.003	\\
	&	2456819.81660	&	0.998	&	0.003	\\
	&	2456819.81724	&	0.996	&	0.003	\\
	&	…	&	…	&	…	\\
\hline							
436	&	2456861.74956	&	1.003	&	0.016	\\
	&	2456861.75085	&	0.999	&	0.015	\\
	&	2456861.75149	&	0.999	&	0.015	\\
	&	2456861.75213	&	1.002	&	0.015	\\
	&	2456861.75277	&	0.999	&	0.015	\\
	&	…	&	…	&	…	\\
\hline
... & ...             & ...    & ...        \\
\hline
\end{tabular}
\end{center}
{Notes: The full table is available in machine-readable form.}
\end{table*}

\section{Light Curve Analysis}
\label{sec:LCanalysis}
In order to find the best fit light curves and planetary parameter of HAT-P-37b, we use the \texttt{TransitFit}, a python package for fitting multi filter and epoch for exoplanet transit observations, which employs the model transit by {\tt\string batman} \citep{kreidberg2015}, and use the dynamic nested sampling routines from {\tt\string dynesty} \citep{speagle2020} to determine the parameters \citep{hayes2021}. 

All 30 light curves were modeled by \texttt{TransitFit} simultaneously. We performed \texttt{TransitFit} by using the nested sampling algorithm with 2000 number of live points and 10 slices sampling. Each transit light curve was individually detrended using the 2nd order polynomial detrending function in \texttt{TransitFit} during the retrieval. The normalized light curves with their uncertainties are available in a machine-readable form in Table~\ref{tab:2}. The initial values of each parameters: orbital period $P$, epoch of mid-transit ${T}_{0}$ (BJD), orbital inclination $i$ (deg), semimajor axis $a$ (in unit of stellar radius, $R_\ast$), the planet's radius $R_p$ (in unit of stellar radius, $R_\ast$) for each filter, are given in Table \ref{tab:initialpara}. The HAT-P-37b's orbit is assumed to be a circular orbit.

\begin{figure*}[htb]
\centering
  \begin{tabular}{cc}
    \includegraphics[scale=0.6, trim= {0 3cm 0 0cm}] {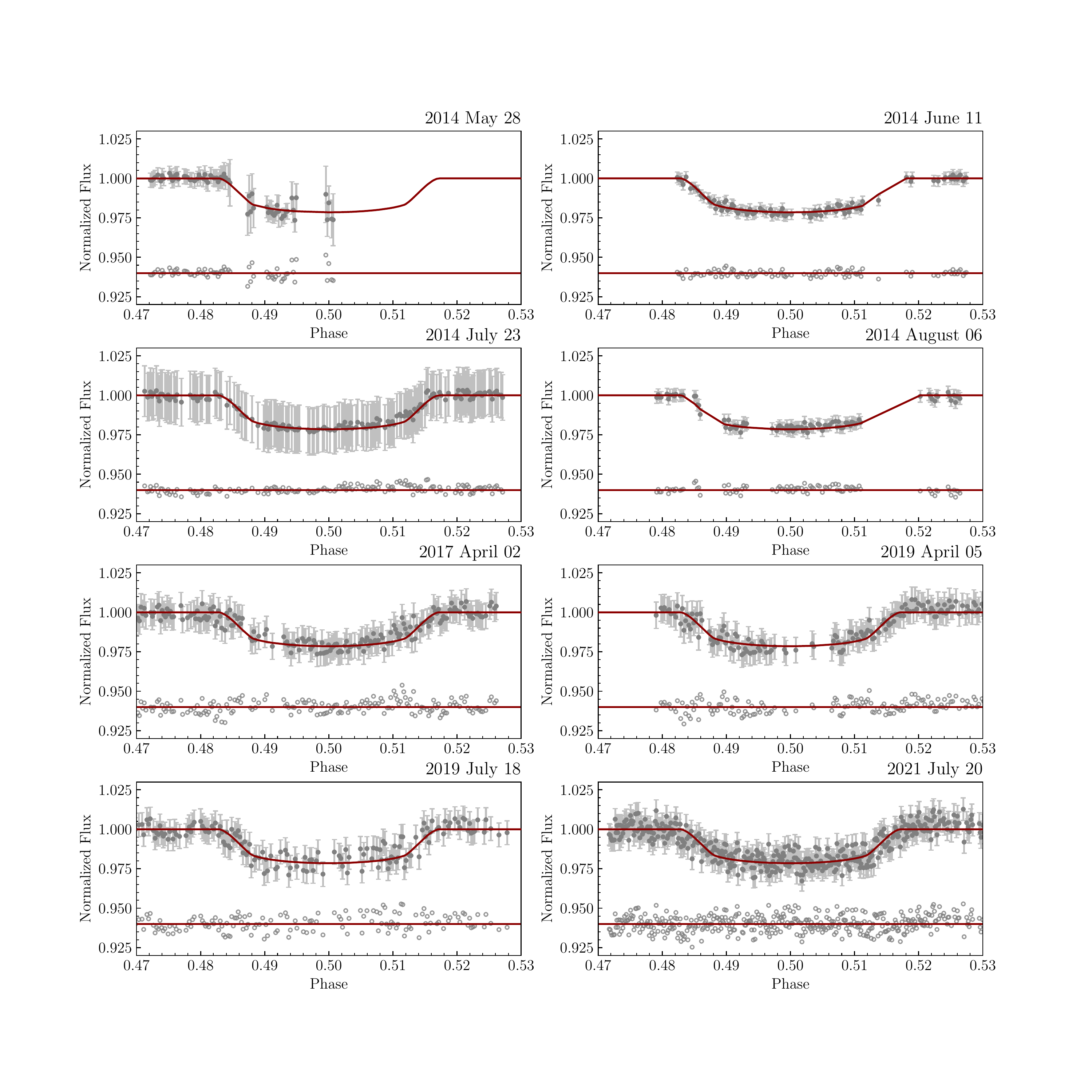} \\
    \includegraphics[scale=0.55]{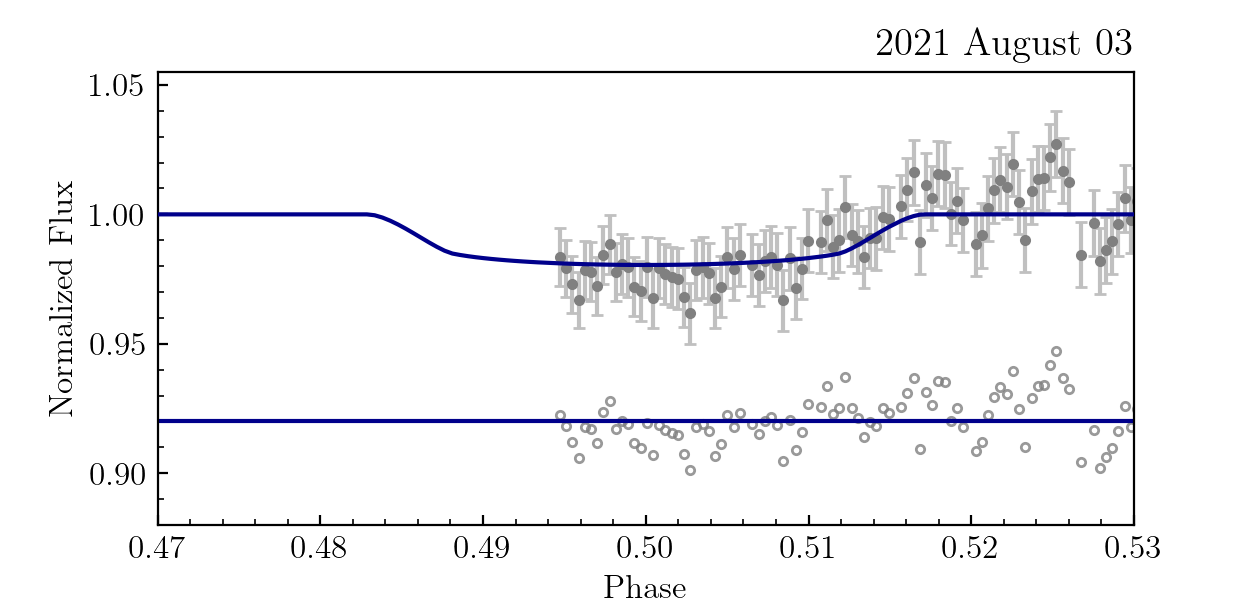} &
    \end{tabular}
    \caption{Normalized Flux as a function of phased-folded transit light curves of HAT-P-37b from our observations. The red and blue lines represent the best-fitting light curves model for $R$-band and $B$-band, respectively. The corresponding residuals with off-sets 0.94 for $R$-band and 0.92 for $B$-band are shown below the light curves.}
    \label{fig:ldfit}
\end{figure*}

In order to obtain the best fits for all light curves, we first used the Uniform distribution to determine the best value of orbital period $P$. A uniform distribution between 2.79738 and 2.79748 days was calculated to provide the best value of orbital period of 2.797434 $\pm$ 4 $\times$ 10$^{-7}$ days. Next, we investigated the existence of TTVs. We used the ability of \texttt{TransitFit} to account for TTVs analysis, by using \texttt{allow\_TTV} function and the best fit period value from the first procedure was fixed, in order to find the mid-transit time ($T_m$) for each epoch. The light curves of HAT-P-37b was phase-folded to each mid-transit time at phase of 0.5 with their best fit models and residuals are shown in Figure \ref{fig:ldfit}. The derived planetary parameters for HAT-P-37b are shown in Table \ref{tab:outpara}. The mid-transit time ($T_m$) for each transit event and corresponding epochs ($E$) are given in Table \ref{tab:midtransit} and discussed in Section \ref{sec:ttv}.

From the analyses, HAT-P-37b has an orbital period of $2.7974341 \pm 4\times10^{-7}$ days with the inclination of $i = 87.0 \pm 0.13$ deg at the star-planet separation of 9.53 $\pm$ 0.1 $R_\ast$. The obtained parameters are compatible with the values from previous studies: \citet{bakos2012,Maciejewski2016,turner2017,yang2021}. However, the {$R_p$/$R_\ast$} value in $B$-band from the fitting is larger than the value analyzed by \citet{turner2017} ( {$R_p$/$R_\ast$} = 0.1253 $\pm$ 0.0021), by about 0.007 $\pm$ 0.002. 

For the analysis of limb-darkening Coefficients (LDCs) of each filter, the \texttt{Coupled} fitting mode in \texttt{TransitFit} is used. The limb-darkening coefficients of each filter is fitted as a free parameter and coupled across wavelengths simultaneously by using the quadratic limb-darkening model and the Limb Darkening Toolkit (LDTk, \citet{husser2013,parviainen2015}). The prior of host star information: stellar effective temperature ${T}_{eff}$ = 5,500 $\pm$ 100 K, metallicity [Fe/H] = 0.03 $\pm$ 0.1 \citep{bonomo2017} and ${log}_{g}$ = 4.54 $\pm$ 0.1 \citep{stassun2019} are adopted during LDCs calculation. The values of limb-darkening coefficients for different filter from coupled LDCs fitting mode are given in Table \ref{tab:limdark}.

\begin{table*}
\begin{center}
\caption{The initial parameters setting and priors used for model the planetary parameters with \texttt{TransitFit}.}
\label{tab:initialpara}          
\small\addtolength{\tabcolsep}{-2pt}
\begin{tabular}{lcc}
\toprule
Parameter  &  Priors & Prior distribution \\
\hline
$P$ [days]         &  2.797434  $^a$       &  Fixed      \\
${T}_{0}$ [BJD]    &  2455642.14318 $\pm$ 0.01  &  A Gaussian distribution  \\
$i$ [deg]    &   86.9 $\pm$1  &  A Gaussian distribution  \\
{$a/R_\ast$}       &   9.3$\pm$1    &  A Gaussian distribution  \\
{$R_p$/$R_\ast$} [$B$-band]  &    (0.11, 0.15)    &   Uniform distribution     \\
{$R_p$/$R_\ast$} [$V$-band]  &    (0.11, 0.15)    &   Uniform distribution    \\
{$R_p$/$R_\ast$} [$R$-band]  &    (0.11, 0.15)    &   Uniform distribution    \\
{$R_p$/$R_\ast$} [Gunn-$r$]  &    (0.11, 0.15)    &   Uniform distribution    \\
{$R_p$/$R_\ast$} [$i'$-band]  &    (0.11, 0.15)    &   Uniform distribution    \\
{$R_p$/$R_\ast$} [No-filter]  &   (0.11, 0.15)     &   Uniform distribution    \\
$e$               & 0       & Fixed     \\
\hline
\end{tabular}
\end{center}
{\textbf{Notes.} The priors of $P$, ${T}_{0}$, ${\it i}$ and {a/R$_\ast$} are set as the values in \citet{bakos2012}. \\ $^a$ This period value was calculated from the first procedure by uniform distribution.}
\end{table*}

\begin{table*}
\begin{center}
\caption{The planetary parameters from \texttt{TransitFit}}
\label{tab:outpara}          
\small\addtolength{\tabcolsep}{-2pt}
\begin{tabular}{lc}
\toprule
Parameter  &  Value \\
\hline
$P$ [days]         &  2.7974341 $\pm$ 4 $\times$ 10$^{-7}$ \\
${T}_{0}$ [BJD]    &  2455642.14768 $\pm$ 0.00011 \\
$i$ [deg]          &   87.0 $\pm$	0.13   \\
{$a/R_\ast$}       &   9.53  $\pm$	0.1    \\
{$R_p$/$R_\ast$} [$B$-band]  &   0.1316 $\pm$	0.0010 \\
{$R_p$/$R_\ast$} [$V$-band]  &   0.1390 $\pm$	0.0006 \\
{$R_p$/$R_\ast$} [$R$-band]  &   0.1380 $\pm$	0.0005 \\
{$R_p$/$R_\ast$} [Gunn-$r$]  &   0.1356 $\pm$	0.0007 \\
{$R_p$/$R_\ast$} [$i'$-band]  &   0.1374 $\pm$	0.0005 \\
{$R_p$/$R_\ast$} [No-filter] &   0.1404 $\pm$	0.0009 \\
\hline
\end{tabular}
\end{center}
\end{table*}

\begin{table*}
\begin{center}
\caption {The limb-darkening coefficients of HAT-P-37b from \texttt{TransitFit} using coupled LDCs fitting mode.}
\label{tab:limdark}
\begin{tabular}{lcc}
\hline
\hline
Filter & $u_0$ & $u_1$\\
\hline
$B$	&	0.441	$\pm$	0.009	&	0.14	$\pm$	0.01	\\
$V$	&	0.448	$\pm$	0.009	&	0.135	$\pm$	0.009	\\
$R$	&	0.445	$\pm$	0.009	&	0.140	$\pm$	0.009	\\
Gunn-$r$	&	0.438	$\pm$	0.009	&	0.15	$\pm$	0.01	\\
$i'$	&	0.446	$\pm$	0.009	&	0.14	$\pm$	0.01	\\
No-Filter	&	0.439	$\pm$	0.009	&	0.15	$\pm$	0.01	\\
\hline									
\end{tabular}\\
\end{center}
\end{table*}

\section{Transit timing analysis}  
\label{sec:ttv}
\subsection{Timing variation models}
\label{subsec:timing model}
In order to perform the timing analyses, mid-transit times of light curves with full transit coverage in 29 epoch listed in Table \ref{tab:midtransit} are considered. The procedure of timing analyses from \citet{patra2017} are followed. The mid-transit times are fitted by three different models: linear ephemeris model, orbital decay model and apsidal precession model, using the \texttt{emcee} Markov Chain Monte Carlo (MCMC) method \citep{foreman2013}. For each model, 50 chains and $10^{5}$ MCMC steps are computed. As mid-transit times are globally obtained from different telescopes for a decade, some data are obtained with precise timing from the GPS clock (e.g. TRT-SRO) while the other synchronized via internet clocks (e.g. P60, MTM-500). The timing error from the clock is less than $1$ s, which is much smaller than the obtained mid-transit time uncertainty. However, the 
uncertainty of mid-transit time could be slightly under-estimated from the fitting. In order to correct the under-estimation, a smoothing constant, $f$, is used to calculate the likelihood as

\begin{equation} \label{eq:lnlike}
\mathrm{ln} \, \mathcal{L} (D|M) = -\frac{1}{2} \left [
\sum_{n}^{N} \mathrm{ln}(2\pi\sigma^{2}) + \chi^2 \right ] \ .
\end{equation}
The $\chi^2$ function is calculated by 
\begin{equation} \label{eq:lnlike_chi}
\chi^2 = \sum_{n}^{N_{LC}}\frac{(D_{\mathrm {n}} - M_{\mathrm{n} } )^2}{s^{2}_n} \ ,
\end{equation}
where $D$ is the observed flux density, $M$ is the modeled flux density and $s^{2}$ is the variance of flux measurement,
\begin{equation}
s^{2}_n = \sigma^2_n + f^{2}M_{\mathrm{n}} \ ,
\end{equation}
$\sigma_n$ is the timing error for the observation.

Firstly, the timing data are fitted with the linear ephemeris model, a constant-period model, as:
\begin{equation}
t(E) =  T_{0,l} + E \times P_{l} \ ,
\end{equation}
where $T_{0,l}$ and $P_{l}$ are the reference time and the orbital period of the linear ephemeris model, respectively. $E$ is the epoch number. Epoch=0 is the transit on 2011 March 21. $t(E)$ is the calculated mid-transit time at a given epoch $E$.

\begin{figure*}
\begin{center}
\includegraphics[width=0.65\textwidth,page=1]{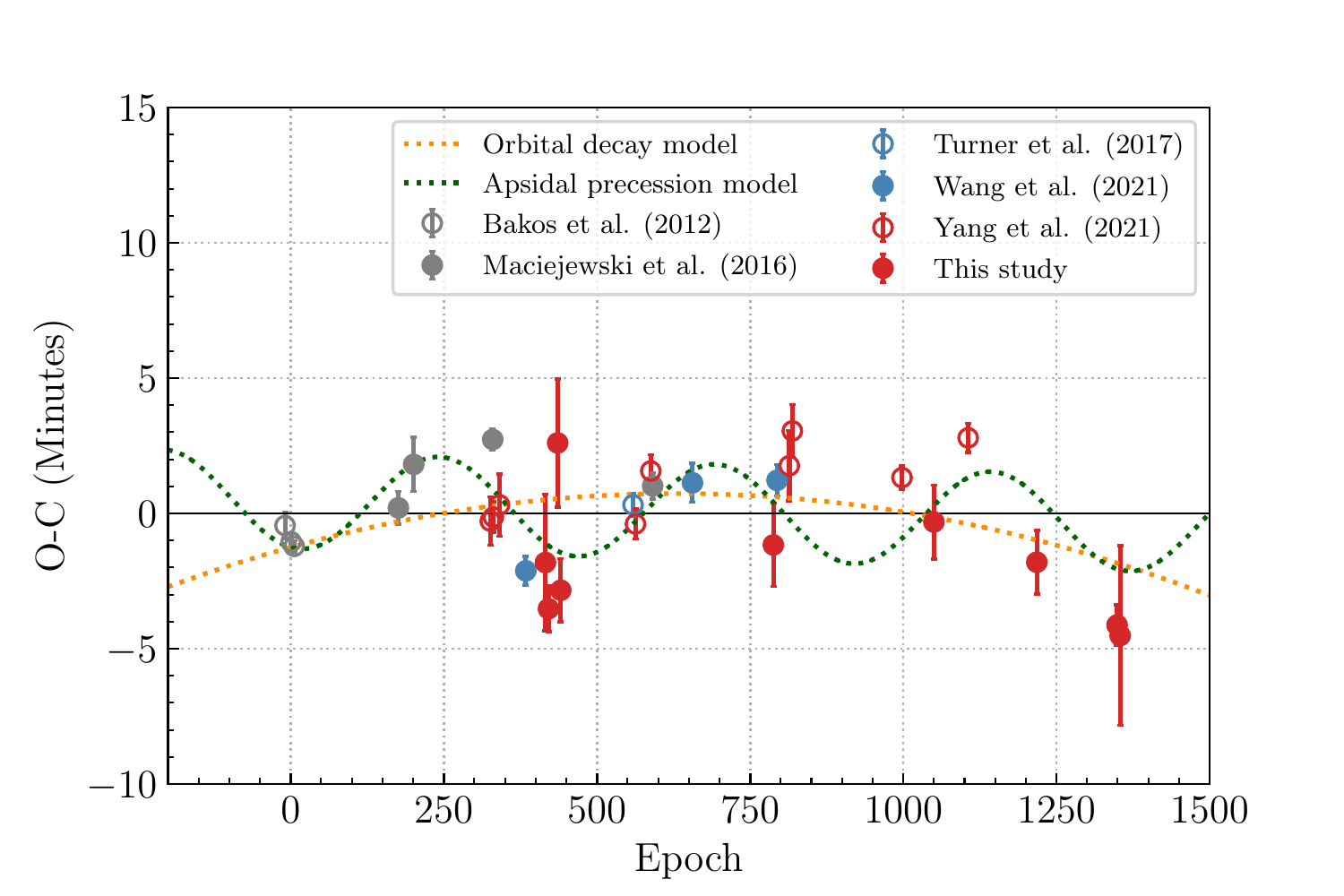}
\caption{O-C diagram and best fitting models for HAT-P-37b with the data from \citet{bakos2012} (Gray unfilled circle), \citet{Maciejewski2016} (Gray filled circle), \citet{turner2017} (Blue unfilled circle), \citet{wang2021} (Blue filled circle), \citet{yang2021} (Red unfilled circle) and this work (Red filled circle). The orange and green dotted lines present the timing residuals of orbital decay and apsidal precession models, respectively.}
\label{fig:oc}
\end{center}
\end{figure*}

The corner plot of MCMC posterior probability distribution of the parameters of constant-period model is shown in Figure \ref{fig:linearMCMC}. The posterior probability distribution provides the best-fit values of $T_{0,l}$ of $2455642.14409^{+0.00046}_{-0.00047}$ (BJD) and $P_{l}$ of $2.797440 ^{+0.000001}_{-0.000001}$ days. The reduce chi-square ${\chi}^{2}_{red}$ of this model is 9.03 with the degree of freedom 27. Using new ephemeris, we constructed the $O-C$ diagram, which is the timing residuals from the difference between the timing data and the best fitting of constant-period model as a function of epoch $E$ as shown in Figure \ref{fig:oc}.

The $O-C$ diagram shows presence of an inverted parabolic with sinusoidal variation trend. Therefore, the orbital decay and apsidal precession models are adopted in order to describe the inverted parabolic trend. For the orbital decay model, we assumed that the orbital period is changing at a steady rate as: 
\begin{equation}
t(E) = T_{0,d} + E \times P_{d} + \frac{1}{2}\frac{\textup{d} P_{d}}{\textup{d} E} E^2 \ ,
\end{equation}
where $T_{0,l}$ is a reference time of the orbital decay model. $P_{d}$ is planetary orbital period of the orbital decay model and $\textup{d}P_{d}/\textup{d}E$ is the change of orbital in each orbit. 

The best-fitting model is shown in Table \ref{tab:timing} with MCMC posterior probability distribution shown in Figure \ref{fig:decayMCMC}. Using the best fitted parameters of this model, the timing residuals as function of epoch $E$ of orbital decay model that can be obtained by subtracting with the best fitting constant-period model is shown in Figure \ref{fig:oc}. The model shows the change of orbital of $\textup{d}P_{d}/\textup{d}E$ = $-7^{+3}_{-3}\times10^{-9}$ days/orbit or $-0.08 ^{+0.03}_{-0.03}$ second per year. The ${\chi}^{2}_{red}$ of the model is 6.69 with the degree of freedom 26. 

The stellar tidal quality factor ${Q}_{\star }^{{\prime}}$ can be expressed as \citep{macie2018}:
\begin{equation}
{Q}_{\star}^{{\prime}} = -\frac{27}{2}\pi\left (\frac{M_p}{M_\star}\right ) \left (\frac{a}{R_\star} \right )^{-5} \left (\frac{dP_d}{dE} \right)^{-1}P, 
\end{equation}
where $M_p$ is the planet mass and $M_{\star}$ is the stellar mass. The values of $M_p$ and $M_{\star}$ are taken from \citet{bakos2012}. Using the value of $\textup{d}P_{d}/\textup{d}E$ from the model fitting, we obtained an estimated value of ${Q}_{\star }^{{\prime}}$ = 250 $\pm$ 10 which is much smaller than the values supported by theoretical models. Therefore, the orbital decay model is unlikely to be a possible choice here. 

The apsidal precession model which can be used to describe the inverted parabolic trend is also adopted. The planet is assume to have a slightly eccentricity $e$ with the argument of pericenter $\omega$ that uniformly precess. The precession model from \citet{gim1995} is used:
\begin{equation}
t(E) = T_{0,a} + E \times P_{a} - \frac{eP_{a}}{\pi}cos\omega(E) \ ,
\end{equation}
where
\begin{equation}
\omega(E) = \omega_{0} + \frac{d\omega}{dE}E \ ,
\end{equation}
\begin{equation}                                                
P_{s} = P_{a}({1-\frac{1}{2\pi}\frac{d\omega}{dE}}) \ .
\end{equation}
$T_{0,a}$ is the reference time of the apsidal precession model. $e$ is the eccentricity, $P_{a}$ is the sidereal period, $\omega$ is the argument of pericenter and $P_{s}$ is the anomalistic period. 

In Table \ref{tab:timing}, the best-fitting parameters from the MCMC posterior probability distribution (Figure \ref{fig:apsiMCMC}) are shown. From the result, a nearly circular orbit ($e$ = $0.0013 ^{+0.0005}_{-0.0004}$) with $\omega_{0}$ = $-0.30 ^{+0.51}_{-0.65}$ rad and $d\omega/dE$ = $0.0143 ^{+0.0009}_{-0.0007}$ rad epoch$^{-1}$ is obtained. The model has ${\chi}^{2}_{red}$ = 4.77 with the degree of freedom 24. As a high precession rate, $d\omega/dE$, is obtained, the timing residual in Figure \ref{fig:oc} shows a sinusoidal trend instead of predicted invert parabolic trend. 

From the results of three fitting models, the apsidal procession model provides the highest maximum log likelihood with $\ln\mathcal{L}= 183$. The linear and orbital decay models provide the lower maximum log likelihoods of 177 and 180, respectively. Comparing the reduced chi-squared of those three best fit models, the apsidal procession model also provides the lowest reduced chi-squared value (${\chi}^{2}_{red}$), which can be supported that the timing variation of HAT-P-37b can be favored by the uniformly precession model. In order to confirm the argument, the Bayesian information criterion (BIC) of those three model are calculated:             
\begin{equation}
BIC = \chi^{2} + k \ln n \ ,
\end{equation}
where $k$ is the number of free parameters, and $n$ is the number of data points. 

From data of 29 epochs ($n$ = 29), the values of BIC from linear, orbital decay and apsidal precession model fits are 250.52, 184.13 and 131.26, respectively. The difference between BIC value of apsidal precession and orbital decay models is $\Delta$BIC = 52.87. Therefore, the apsidal precession model is favourable for the timing data fitting.

However, the apsidal precession model fitting shows the sinusoidal variation with transit time data. This variation might be affected by light-time effect (LiTE) due to the third component in the HAT-P-37 system. Therefore, the timing variation due to a third body in the system is analysed in the rest of this section.

\begin{table*}
\begin{center}
\caption {Mid-transit times ($T_{m}$) and timing residuals ($O-C$) for HAT-P-37b from 17 transit light curves derived from \texttt{TransitFit}. Epoch=0 is the transit on 2011 March 21.}
\label{tab:midtransit}
\begin{tabular}{lccc}
\hline
\hline
Epoch & $T_{m} +2450000$ & $O-C$ & Data Sources \\
      & [$BJD_{TDB}$]    &  [days] &            \\
\hline
-9	&	5616.96681	$\pm$	0.00033	&	-0.00031	&	(a)	\\
1	&	5644.94080	$\pm$	0.00018	&	-0.00073	&	(a)	\\
6	&	5658.92788	$\pm$	0.00014	&	-0.00085	&	(a)	\\
176	&	6134.49375	$\pm$	0.00042	&	0.00014	&	(b)	\\
201	&	6204.43087	$\pm$	0.00069	&	0.00126	&	(b)	\\
326	&	6554.10947	$\pm$	0.00062	&	-0.0002	&	(e)	\\
330	&	6565.30133	$\pm$	0.00028	&	0.0019	&	(b)	\\
331	&	6568.09678	$\pm$	0.00039	&	-0.00009	&	(e)	\\
341	&	6596.07149	$\pm$	0.00080	&	0.00022	&	(e)	\\
384	&	6716.35974	$\pm$	0.00037	&	-0.00147	&	(d)	\\
416	&	6805.87805	$\pm$	0.00176	&	-0.00126	&	(f)	\\
421	&	6819.86407	$\pm$	0.00058	&	-0.00245	&	(f)	\\
436	&	6861.82993	$\pm$	0.00164	&	0.00181	&	(f)	\\
441	&	6875.81335	$\pm$	0.00081	&	-0.00197	&	(f)	\\
559	&	7205.91352	$\pm$	0.00030	&	0.00022	&	(c)\textbf{$^\star$}	\\
563	&	7217.10278	$\pm$	0.00039	&	-0.00027	&	(e)	\\
588	&	7287.04016	$\pm$	0.00041	&	0.00109	&	(e)	\\
591	&	7295.43210	$\pm$	0.00033	&	0.00071	&	(b)	\\
656	&	7477.26580	$\pm$	0.00050	&	0.00079	&	(d)	\\
788	&	7846.52635	$\pm$	0.00107	&	-0.00081	&	(f)	\\
794	&	7863.31265	$\pm$	0.00040	&	0.00085	&	(d)	\\
814	&	7919.26183	$\pm$	0.00090	&	0.00122	&	(e)	\\
819	&	7933.24993	$\pm$	0.00068	&	0.00212	&	(e)	\\
998	&	8433.99057	$\pm$	0.00030	&	0.00092	&	(e)	\\
1050	&	8579.45633	$\pm$	0.00095	&	-0.00022	&	(f)	\\
1106	&	8736.11516	$\pm$	0.00037	&	0.00194	&	(e)	\\
1218	&	9049.42530	$\pm$	0.00081	&	-0.00125	&	(f)	\\
1349	&	9415.88838	$\pm$	0.00052	&	-0.00286	&	(f)	\\
1354	&	9429.87531	$\pm$	0.00231	&	-0.00313	&	(f)	\\
\hline
\end{tabular}\\
\end{center}
{\textbf{Note.} Data Source: (a) \citet{bakos2012} (b) \citet{Maciejewski2016} (c) \citet{turner2017}, (d) \citet{wang2021}, (e) \citet{yang2021} and (f) this study;
$^\star$ : $T_m$ of this epoch was averaged from two transit light curves.}
\end{table*}

\begin{table*}
\begin{center}
\caption {Priors with uniform distribution and best fitting parameters from MCMC transit timing analyses.}
\label{tab:timing}
\begin{tabular}{lcc}
\hline
\hline
Parameter & Uniform distribution priors & Best fit values\\
\hline
\textbf{Constant-period Model}  &                &	              \\
$P_{orb,l}$ [days]                &  (2.7, 2.9)    & $2.797440 ^{+0.000001}_{-0.000001}$     \\
$T_{0,l}$ [$BJD_{TDB}$]         &  (2455642.141, 2455642.148)   &   $2455642.14409 ^{+0.00046}_{-0.00047}$  \\
\textbf{Orbital Decay Model}    &        &	    \\
$P_{orb,d}$ [days]                &  (2.7, 2.9)    &   $2.797445 ^{+0.000002}_{-0.000002}$     \\
$T_{0,d}$ [$BJD_{TDB}$]         &  (2455642.141, 2455642.148)   &  $2455642.14322 ^{+0.00058}_{-0.00057}$  \\
$dP/dE$ [days/orbit]            &  (-0.5, 0.5) &   $-7^{+3}_{-3}\times10^{-9}$   \\
\textbf{Apsidal Precession Model}   &        &	      \\
$P_{s}$ [days]                   &  (2.7, 2.9)      &  $2.797440 ^{+0.000001}_{-0.000001}$      \\
$T_{0,a}$ [$BJD_{TDB}$]          &  (2455642.142, 2455642.148)  &  $2455642.14436 ^{+0.00042}_{-0.00040}$  \\
$e$                              &  (0, 0.003)         &   $0.0013 ^{+0.0005}_{-0.0004}$    \\
$\omega_{0}$ [rad]               &  (-$\pi$, $\pi$)    &   $-0.30 ^{+0.51}_{-0.65}$        \\
$d\omega/dE$ [rad epoch$^{-1}$]  &  (0, 0.025)         &  $0.0143 ^{+0.0009}_{-0.0007}$    \\
\hline
\end{tabular}\\
\end{center}           
\end{table*}

\subsection{The frequency analysis of TTVs}
\label{subsec:gls}
\begin{figure*}[htb]
\centering
  \begin{tabular}{cc}
    \includegraphics[width=0.65\textwidth,page=1]{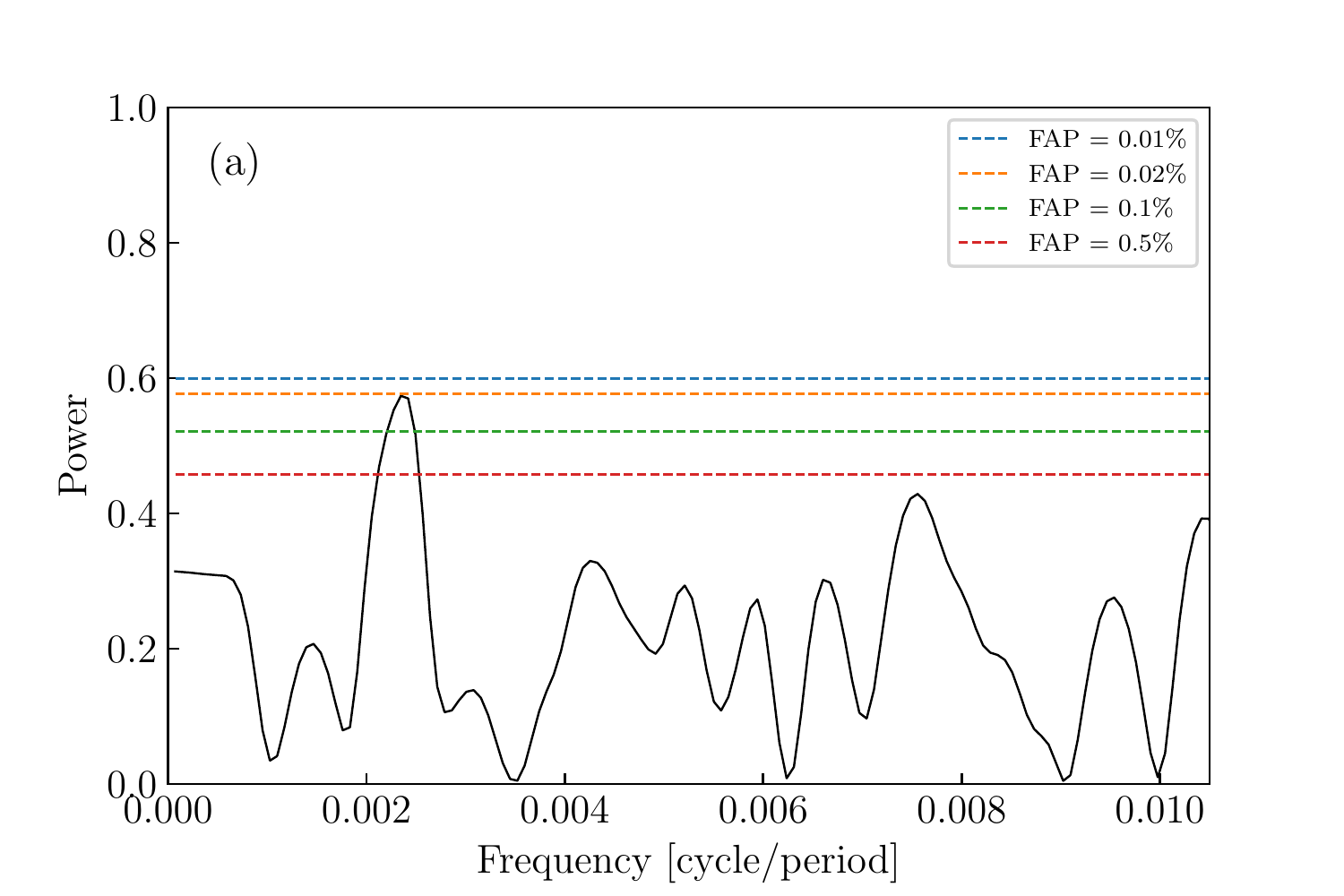} \\
    \includegraphics[width=0.65\textwidth,page=1, trim ={0 0 0 0.3cm}]{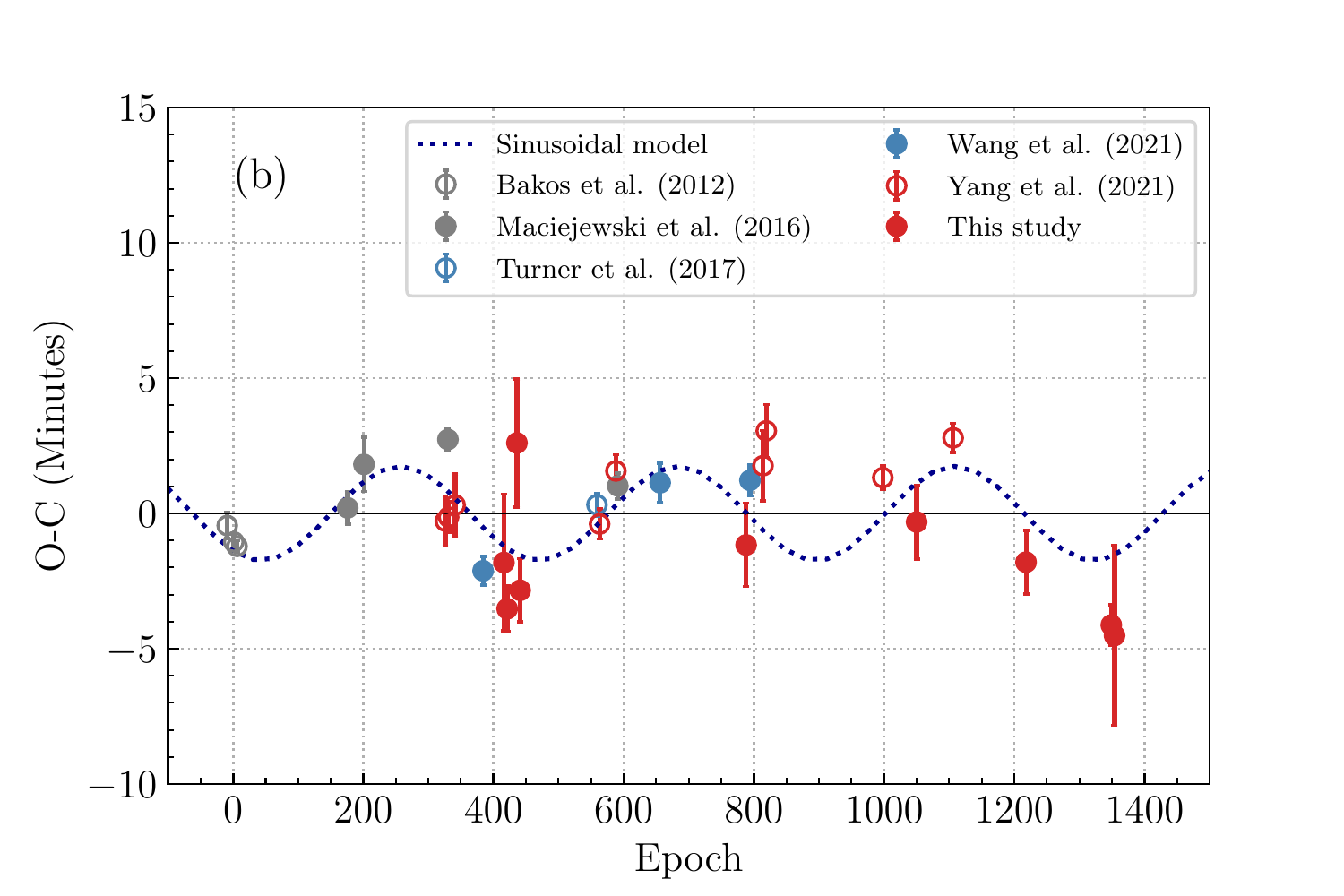} &
    \end{tabular}
    \caption{Searching for possible periodicity of TTVs of HAT-P-37b (a) GLS periodogram for timing residuals from Table \ref{tab:midtransit}. The dashed lines indicate the FAP levels. (b) O-C diagram and the best-fit of sinusoidal variability from the frequency of the highest power peak, FAP = 0.02\% (blue dotted line).}
    \label{fig:gls}
\end{figure*}

In order to investigate the sinusoidal of TTVs on HAT-P-37b data. The Generalized Lomb-Scargle periodogram (GLS; \citet{zech2009}) in {\tt\string PyAstronomy}\footnote{PyAstronomy: \texttt{https://github.com/sczesla/PyAstronomy}} routines \citep{pya2019} is used to search for periodicity in the timing residuals (O-C) data given in Table \ref{tab:timing}. 
The False Alarm Probability (FAP) is calculated in order to provide the probability of peak detection 
from the highest power peak. 
The result of GLS is shown in periodogram of the power spectrum as a function of frequency in Figure \ref{fig:gls}. In the periodogram, the highest power peak = 0.574 at frequency of 0.0023 $\pm$ 0.0001 cycle$/$period (epoch) calculated from FAP of 0.02 $\%$ is found. However, this FAP level consists with noise and no significant of periodicity. The FAP levels of 0.5$\%$, 0.1$\%$ and 0.01$\%$ are presented. 

Nevertheless, the frequency of the highest power peak is tested by assuming TTVs with a sinusoidal variability. We apply the procedure described in \citet{von2019}. The timing residuals were fitted through a fitting function as:  
\begin{equation}
\label{eq:frequency}
TTVs(E) = A_{TTVs}\sin(2{\pi}fE - \phi),
\end{equation}
where $A_{TTVs}$ is amplitude (minutes) of the timing residuals, $f$ is the frequency on the highest peak of power periodogram, and $\phi$ is phase. 

From the fitting, $A_{TTVs}$ = 1.74 $\pm$ 0.17 minutes and $\phi$ = 2.2 $\pm$ 0.08 with the best fitted ${\chi}^{2}_{red}$ = 4.39 and BIC = 125.25 are obtained. The timing residuals with the best-fit of sinusoidal variability is plotted in Figure \ref{fig:gls}. The reduced chi-squared and BIC values of the sinusoidal model is lower than the values from the apsidal precession model. Therefore, there is a possibility to have an additional exoplanet in the system.

An additional exoplanet that has orbital period near the the first-order resonance of HAT-P-37b with a co-planar orbit is assumed. With a first order mean-motion resonance, \textit{j:j-1}, the perturber planet mass can be calculated from \citet{lith2012} equation:
\begin{equation}
    V = P\frac{\mu^{'}}{\pi j^{2/3} (j-1)^{1/3} \Delta} \left(-f-\frac{3}{2} \frac{Z^{*}_{free}}{\Delta}\right) \ ,
\label{eq:masspert}
\end{equation}
where $V$ is the amplitude of transit time variation. For our case $V$ = 1.74 minute. \textit{P} is period of HAT-P-37b, $\mu^{'}$ is the outer planet mass, $\Delta$ is the normalized distance to resonance, $f$ is sums of Laplace coefficients with order-unity values and $Z^{*}_{free}$ is the dynamical quantity that controls the TTV signal (more explanation given in \citet{lith2012}). 

From the calculation, in the case of 1:2 mean motion resonance, the mass of the additional planet could be as small as 0.0002$M_{Jup}$ or 0.06$M_{\oplus}$ with the period of 5.58 days. The perturber mass is lighter than the Earth. Assuming the planet is a rocky planet with the Earth's density, the planet has the radius of 0.4 $R_{\oplus}$. If the planet transit the host star, it will produce the transit depth of $1.75 \times 10^{-5}$, which cannot be detected by the light curve precision in this work. 

\subsection{Upper mass limit for an additional planet}
\begin{figure*}
\begin{center}
\includegraphics[width=0.65\textwidth,page=1]{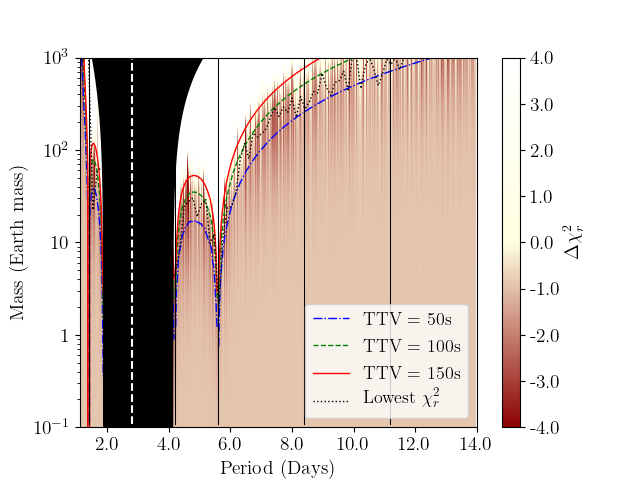}
\caption{Upper mass limit of the perturbing planet in the HAT-P-37 system. The upper mass limit for TTV amplitude of 50,100 and 150 seconds are shown in blue dash-dot, green dashed and red line, respectively. The best $\Delta\chi^2_{red}$ within a 0.05 period ratio bin is presented by black dotted line. The contours represent the best $\Delta\chi^2_{red}$ between the best TTV model and the best linear model. The unstable orbit regions shown in black shaded region. The write vertical dashed line displays the orbital period of HAT-P-37b. The black vertical lines show 2:1 3:2 1:2 1:3 1:4 orbital period resonance from left to right, respectively.}
\label{fig:uppermass}
\end{center}
\end{figure*}

From Section \ref{subsec:gls}, an additional planet orbits near 1:2 mean motion resonance of HAT-P-37b is investigated. In this section the upper mass limit upper mass limit for an additional planet near HAT-P-37b is found from the timing variation. The method of searching for the upper mass limit of the second planet given in \citet{awiphan2016} are followed. Firstly, we assumed that two-planets are coplanar and circular orbits. The unstable regions is calculated from the mutual Hill sphere between HAT-P-37b and the perturber by \citet{fab2012} ;
\begin{equation}
    r_{H} = \frac{a_{in} + a_{out}}{2}\left(\frac{M_{in} + M_{out}}{3M_\star}\right)^{1/3} \ ,
\end{equation}
where $a_{in}$ and $a_{out}$ are the semi-major axis of the inner and outer planets, respectively. The boundary of stable orbit is when the separation of the planets semimajor axes ($a_{out} - a_{in}$) is larger than $2\sqrt{3}$ of the mutual Hill sphere. The region of unstable orbits is shown by black shaded area in Figure \ref{fig:uppermass}.     

The {\tt\string TTVFaster}\footnote{TTVFaster: \texttt{https://github.com/ericagol/TTVFaster}} by \citet{deck2016}; a package for dynamical analysis that is accurate to first order in orbital eccentricity to search for TTVs signal of secondary planet, is used. The period ratio between the perturber planet and HAT-P-37b between 0.3 and 5.0 with 0.01 steps is set with the mass range between $10^{-1}$ and $10^3$ Earth mass in logarithmic scale. From the amplitude signal 104.4 seconds on O-C diagram in Section \ref{subsec:gls}, we calculate the upper mass limits corresponding to TTV signal amplitudes of 50, 100 and 150 seconds shown in Figure \ref{fig:uppermass}. 

Finally, using the comparison between ${\chi}^{2}_{red}$ values of the best linear fitting model, a single-planet model, and ${\chi}^{2}_{red}$ of the signal from the two planets model by {\tt\string TTVFaster} as:
\begin{equation}
    \Delta\chi^2_{red} = \chi^2_{red} - \chi^2_{red,l} \ , 
\end{equation}
where $\chi^2_{red}$ is the best fitting of second planet model (TTV model) at given mass and period and $\chi^2_{red,l}$ is the best fitting of single planet model. $\chi^2_{red,l}$ = 9.03 is obtained from Section \ref{subsec:timing model}. The $\Delta\chi^2_{red}$ is shown as a function of perturber mass and period in Figure \ref{fig:uppermass}. The regions of negative values of $\Delta\chi^2_{red}$ are near the 100 seconds TTV amplitude as predicted. From the result, we can conclude that there is no Saturn mass planet within 1:2 orbital period resonance.
 
\section{HAT-P-37b Atmosphere}
\label{sec:atmosphere}
The investigation of variations in the transit depth with wavelength of HAT-P-37b was discussed by \citet{turner2017}. They found that HAT-P-37b shows a small transit depth in $B$-filter may be caused by the TiO/VO absorption. From this investigation, the study of HAT-P-37b transmission spectrum is considered.

\begin{figure*}
\begin{center}
\includegraphics[width=0.9\textwidth,page=1]{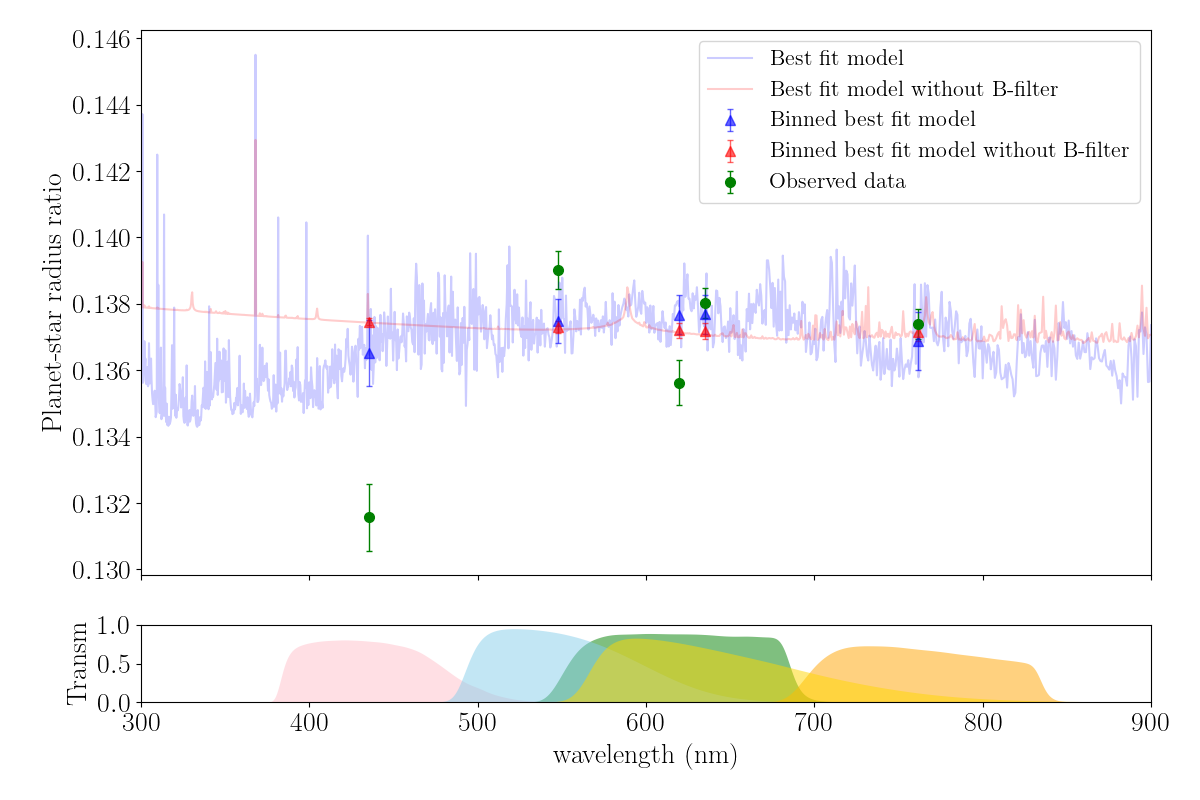}
\caption{The best fit transmission spectrum of HAT-P-37b with synthetic models generated from \texttt{PLATON} retrieval (Top) and the band-pass filters: \textit{B, V, R, r} and \textit{i} band (from left to right) (Bottom). }
\label{fig:atmos}
\end{center}
\end{figure*}

From {$R_p$/$R_\ast$} values from different filters obtained from Section \ref{sec:LCanalysis}, the broad-band transmission spectrum of HAT-P-37b is shown in Figure \ref{fig:atmos}. PLanetary Atmospheric Transmission for Observer Noobs ({\tt\string PLATON} \footnote{PLATON: \texttt{https://github.com/ideasrule/platon}}; \citet{zhang2019}) is used to model and retrieves atmospheric characteristics of the transmission spectrum. For \texttt{PLATON} retrieval run, 1000 number of live points are performed by the nested sampling method with the priors as in Table \ref{tab:atmosphere}.

The transit depths of five filters: $B$, $V$, $R$, $r'$ and $i'$ bands, with wavelength coverage between 390 and 779 nm are obtained in Section \ref{sec:LCanalysis}. From the fitting, the HAT-P-37b radius at 100,000 Pa of 1.11 $R_{Jup}$ and mass of 1.17 $M_{Jup}$ are obtained with the host stellar radius of 0.86 $R_\odot$ (Table \ref{tab:atmosphere} and Figure \ref{fig:atmos_corner_BVRri}). The model shows the temperature of HAT-P-37b atmosphere with the isothermal model of 1,800 K. However, the model provides the ${\chi}^{2}$ value 42 with a large discrepancy in $B$-filter data, which obtained from two transits from \citet{turner2017} and a partial transit from the TRT-SBO. As the model cannot fit the depth in $B$ filter, the depth is excluded in the further analysis. 

Another model without the depth in $B$-fitting is fitted with \texttt{PLATON}. Four transit depth data in $V$, $R$, $r'$ and $i'$ filters with wavelength range 501 to 779 nm are modeled. The best fitting results is shown in Table \ref{tab:atmosphere} and Figure \ref{fig:atmos_corner_VRri}. The result provides the cooler atmospheric temperature of 1,100 K with ${\chi}^{2}$ of 18. However, the cloudy model, i.e. a model with a constant transit depth, of the data without the depth in $B$-filter provides the planet-star radius ratio of 0.137 and ${\chi}^{2}$ of 16, which is lower than the chi-square values of both \texttt{PLATON} fitting models. Therefore, there is a possibility that thick clouds cover the HAT-P-37b atmosphere. 

\begin{table*}
\begin{center}
\caption{The priors and prior distribution used for \texttt{PLATON} retrieval and the best-fit parameters of each wavelength range analysis from the retrieval.}
\label{tab:atmosphere}          
\small\addtolength{\tabcolsep}{-2pt}
\begin{tabular}{lcccc}
\toprule
Parameter  & Priors & Prior distribution & Wavelength range  & Wavelength range   \\
  &   &   & 390-779 nm  & 501-779 nm   \\
\hline
$R_s$ [$R_\odot$]   &  $0.87 \pm 0.06$ &  Gaussian distribution  & $0.86^{+0.02}_{-0.02}$ & $0.86^{+0.02}_{-0.02}$      \\
$M_p$ [$M_{jup}$]   &  $1.169 \pm 0.1$ &  Gaussian distribution  & $1.17^{+0.04}_{-0.04}$ & $1.18^{+0.04}_{-0.04}$      \\
$R_p$ [$R_{jup}$]   &  (1.06, 1.30)  &  Uniform distribution  & $1.11^{+0.03}_{-0.03}$ & $1.15^{+0.03}_{-0.03}$      \\
$T$ [$K$]   &  (635, 1900)       &  Uniform distribution  & $1900^{+400}_{-500}$ & $1100^{+400}_{-400}$      \\
$\log_{\textup{Scattering Factor}}$  &  (0, 2)       &  Uniform distribution  & $0.9^{+0.7}_{-0.6}$ & $1.1^{+0.7}_{-0.7}$      \\
$\log_{Z/Z_{\odot}}$  &   (-0.8, 1.6)      &  Uniform distribution  & $0.6^{+0.8}_{-0.8}$ & $0.3^{+0.8}_{-0.8}$      \\
C/O ratio        &    0.3          &  Fixed & 0.3 & 0.3 \\
Error multiple   &   (0.5, 5)      &  Uniform distribution  & $3.6^{+0.8}_{-0.9}$ & $1.8^{+1.3}_{-0.9}$      \\
\hline
\end{tabular}
\end{center}
{\textbf{Notes.} The priors of $R_s$, $M_p$ and $R_p$ are set as the values from \citet{bakos2012}.}
\end{table*}

\section{Conclusions}
\label{sec:conclusion}
In this work, the photometric observations and studies of a hot Jupiter HAT-P-37b are performed. Nine transit light curves are obtained from three telescopes: 60-inch telescope at Palomar Observatory, the 50-cm Maksutov telescope at the Crimean Astrophysical Observatory, Crimea, and 0.7-m Thai Robotic Telescope at Sierra Remote Observatories. The observational data are combined with 21 published light curves. Using the \texttt{TransitFit}, the HAT-P-37b parameters and the mid-transit times are obtained. From the fitting, the planet has an orbital period of $2.7974341 \pm 4\times10^{-7}$ days, the inclination of $i = 87.0 \pm 0.13$ deg and the star-planet separation of 9.53 $\pm$ 0.1 $R_\ast$ which are consistent with previous works. The planet-star radius ratios in five wavebands: $B$, $V$, $R$, $r'$ and $i'$, are obtained. Nevertheless, the fitted transit depth in $B$-band from this study is larger than the value analyzed by \citet{turner2017}.

From the fitting, 29 mid-transit times are obtained. The O-C diagram of HAT-P-37b mid transit time show an inverted parabolic with sinusoidal variation trend. Therefore, three timing variation models: linear ephemeris model, orbital decay model and apsidal precession model are used to analyse the variation. The stellar tidal quality factor ${Q}_{\star}^{{\prime}}$ is determined to be 250 $\pm$ 10 which is far too small and inconsistent with theoretical estimation. The apsidal precession is favourable for the timing data fitting with $d\omega/dE$ = $0.0143 ^{+0.0009}_{-0.0007}$ rad epoch$^{-1}$, maximum log likelihood of $\ln\mathcal{L}= 183$ and ${\chi}^{2}_{red}$ of 4.77. However, due to the large value of $d\omega/dE$, the model shows the sinusoidal variation on transit time data which might be explained by light-time effect (LiTE) of the third body in the system. Therefore, the timing residuals (O-C) data were considered by frequency analysis and sinusoidal variability model fitting. From the analysis, the TTVs amplitude signal of 1.74 $\pm$ 0.17 minutes is obtained. If the third body orbit is at the 1:2 mean motion resonance, its mass can be as small as 0.0002$M_{Jup}$ or 0.06$M_{\oplus}$. The upper mass limit for the perturber planet in HAT-P-37 system is calculated using the {\tt\string TTVFaster} package. The results shows that there is no nearby ($P < 3$ days) planet with mass heavier than Saturn around HAT-P-37b. The mutual Hill sphere regions between orbital period of 1.9 - 4.2 days represents the excluding of the presence of a nearby planet. 

For the transmission spectroscopy analysis of HAT-P-37b, the transit depths of five filters $B$, $V$, $R$, $r'$ and $i'$ bands with wavelength range between 390 to 779 nm are modelled by \texttt{PLATON} fitting model. The model shows the temperature of HAT-P-37b atmosphere with the isothermal temperature model of 1800 K with a large ${\chi}^{2}$ value (${\chi}^{2}$=42) due to a large discrepancy in $B$-filter data. Therefore, the model without the transit-depth in $B$-filter is considered. The model provide a cooler atmospheric temperature of 1,100 K with ${\chi}^{2}$=18. However, this chi-square value still larger than the value of constant transit depth mode (${\chi}^{2}$=16), which can infer to a cloudy atmospheric model. 

Although, a small additional planet and a cloudy atmosphere model of HAT-P-37b can be concluded from the analyses in this work. Additional high-precision observation data in both transit-timing and transit-depth, especially in blue waveband, is needed before the perturber and the atmospheric model can be confirmed.

\begin{acknowledgments}
We thank the anonymous referee for good comments and suggestions 
which helped to improve the quality of this paper.
This work is supported by the grant from the Ministry of Science and Technology (MOST), Taiwan. The grant numbers are MOST 109-2112-M-007-007 
and MOST 110-2112-M-007-035. There are two nights' observational data based on observations made with the Thai Robotic Telescopes, which are operated by the National Astronomical Research Institute of Thailand (Public Organization). This work is also partially supported by a National Astronomical Research Institute of Thailand (Public Organization) research grant. We would like to thank all the authors of previous HAT-P-37b papers for kindly providing their observational data to make this work possible. 
\end{acknowledgments}
%
\vspace{5mm}
\facilities{P60 (Palomar Observatory), MTM-500 (CrAO) and 0.7-m (TRT-SRO).}
\software{{\tt\string sextractor} \citep{berlin1996}, {\tt\string Astrometry.net} \citep{lang2010}, \texttt{TransitFit} \citep{hayes2021}, {\tt\string TTVFaster} \citep{deck2016} and {\tt\string PLATON} \citep{zhang2019}. }

\appendix
\counterwithin{figure}{section}
\section{Posterior probability distribution for three TTVs models MCMC fitting parameters.}
\begin{figure*}[h]
\begin{center}
\includegraphics[scale=0.5]{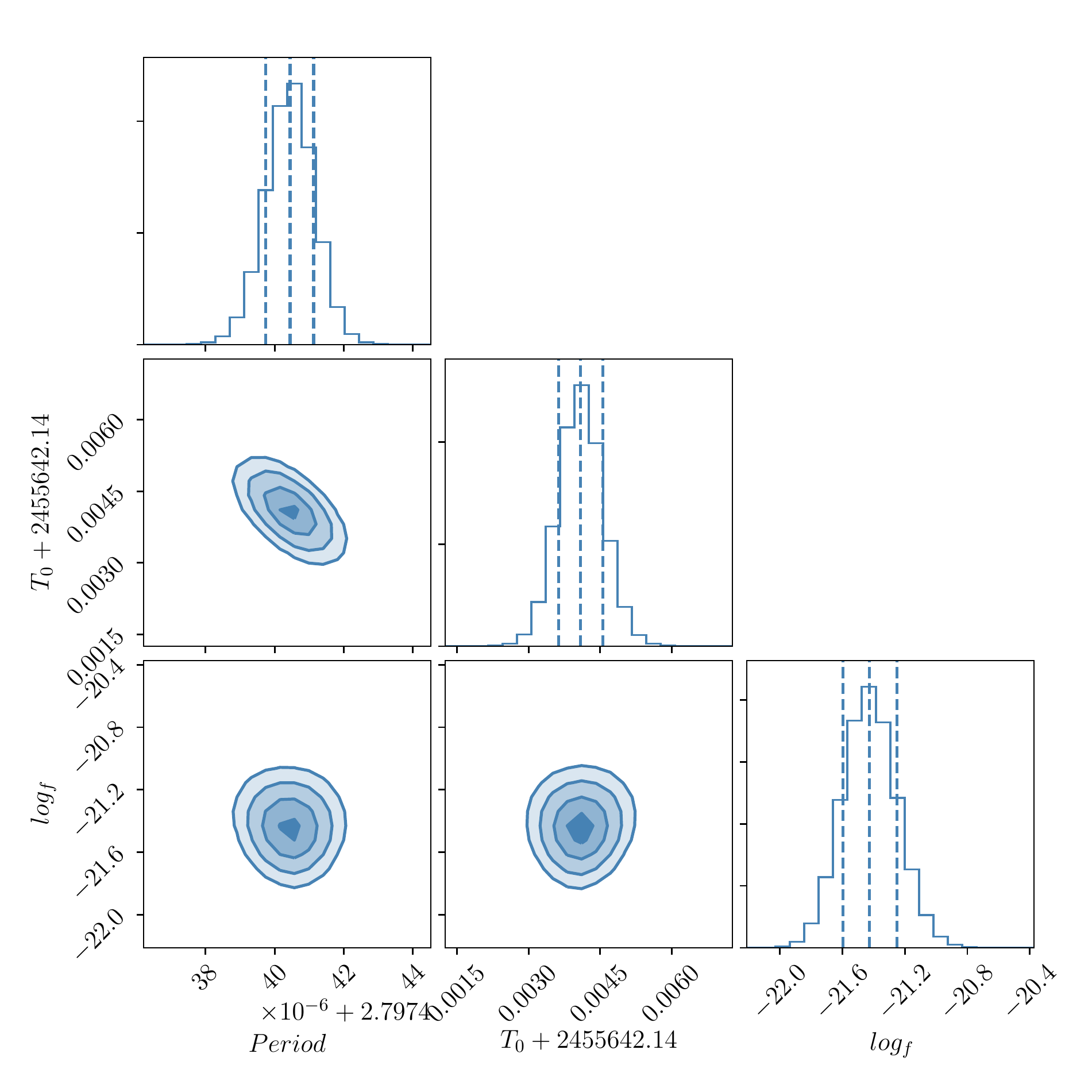}
\caption{Posterior probability distribution of the constant-period model MCMC fitting parameters.}
\label{fig:linearMCMC}
\end{center}
\end{figure*}

\begin{figure*}[h]
\begin{center}
\includegraphics[scale=0.45]{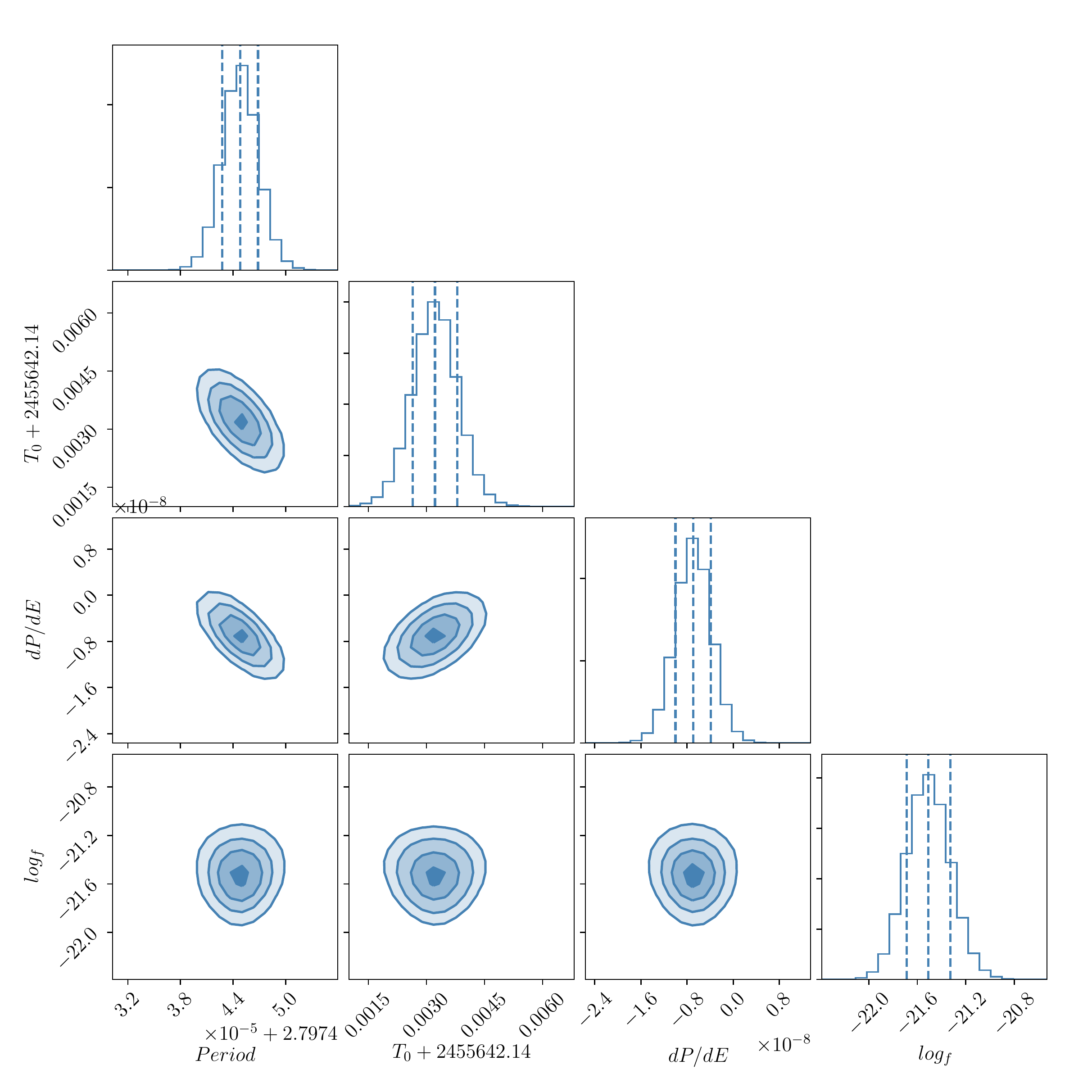}
\caption{Posterior probability distribution of the orbital decay model MCMC fitting parameters.}
\label{fig:decayMCMC}
\end{center}
\end{figure*}

\begin{figure*}[h]
\begin{center}
\includegraphics[scale=0.4]{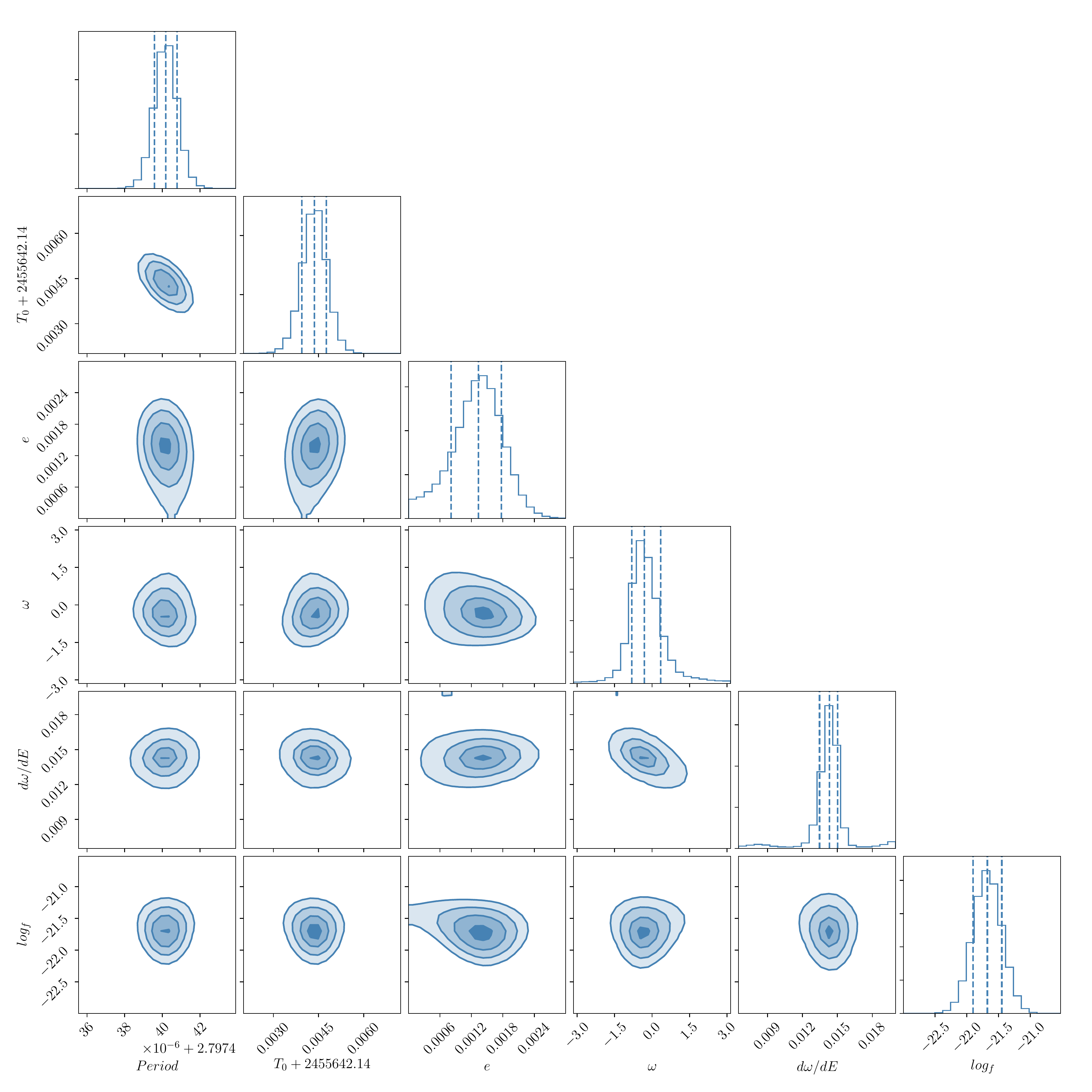}
\caption{Posterior probability distribution of the apsidal precession model MCMC fitting parameters.}
\label{fig:apsiMCMC}
\end{center}
\end{figure*}

\section{Posterior probability distributions from \texttt{PLATON} for HAT-P-37b transmission spectrum study.}
\begin{figure*}[h]
\begin{center}
\includegraphics[scale=0.35]{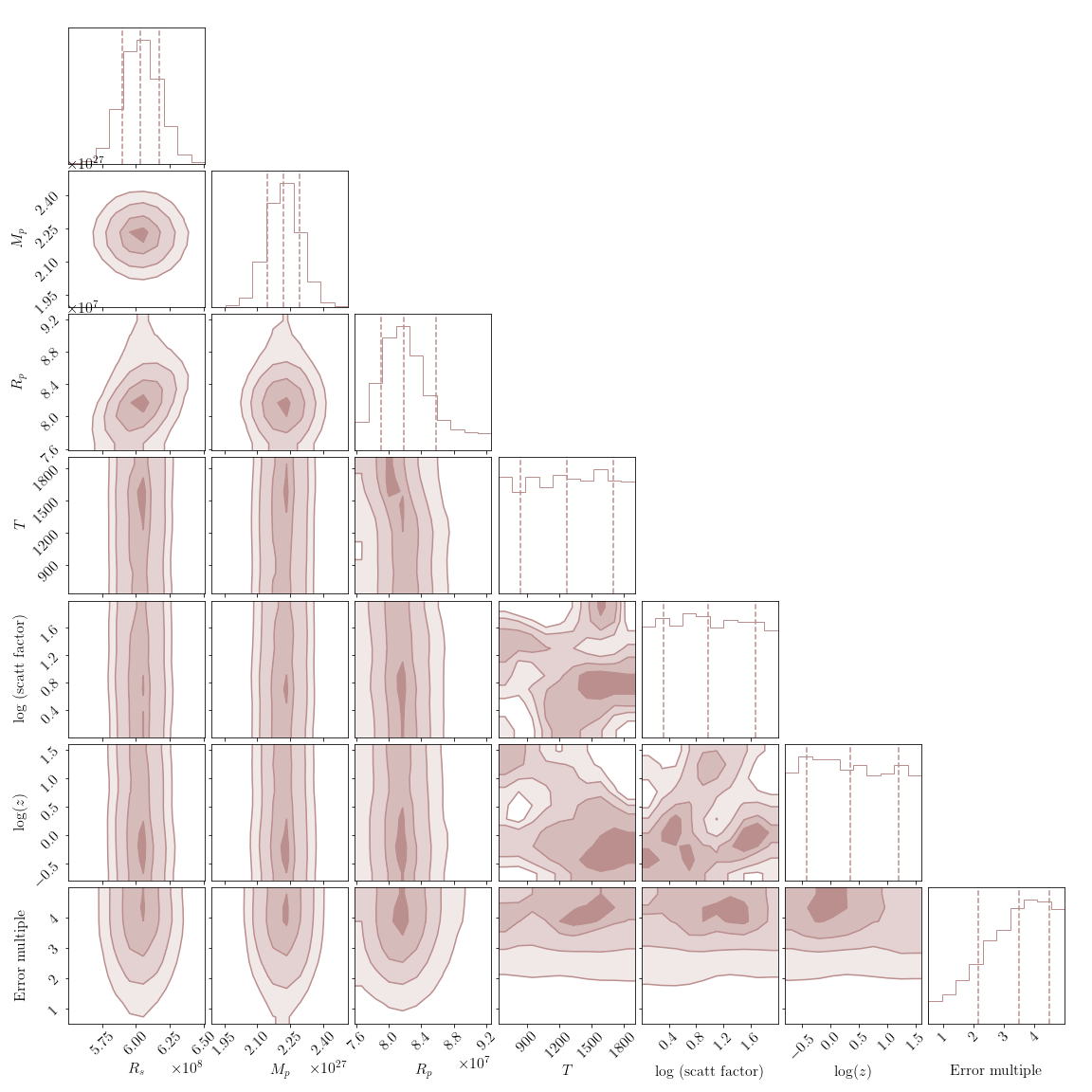}
\caption{Posterior probability distributions from \texttt{PLATON} retrieval for HAT-P-37b of wavelength range 390-779 nm (including $B$-filter).}
\label{fig:atmos_corner_BVRri}
\end{center}
\end{figure*}

\begin{figure*}[h]
\begin{center}
\includegraphics[scale=0.35]{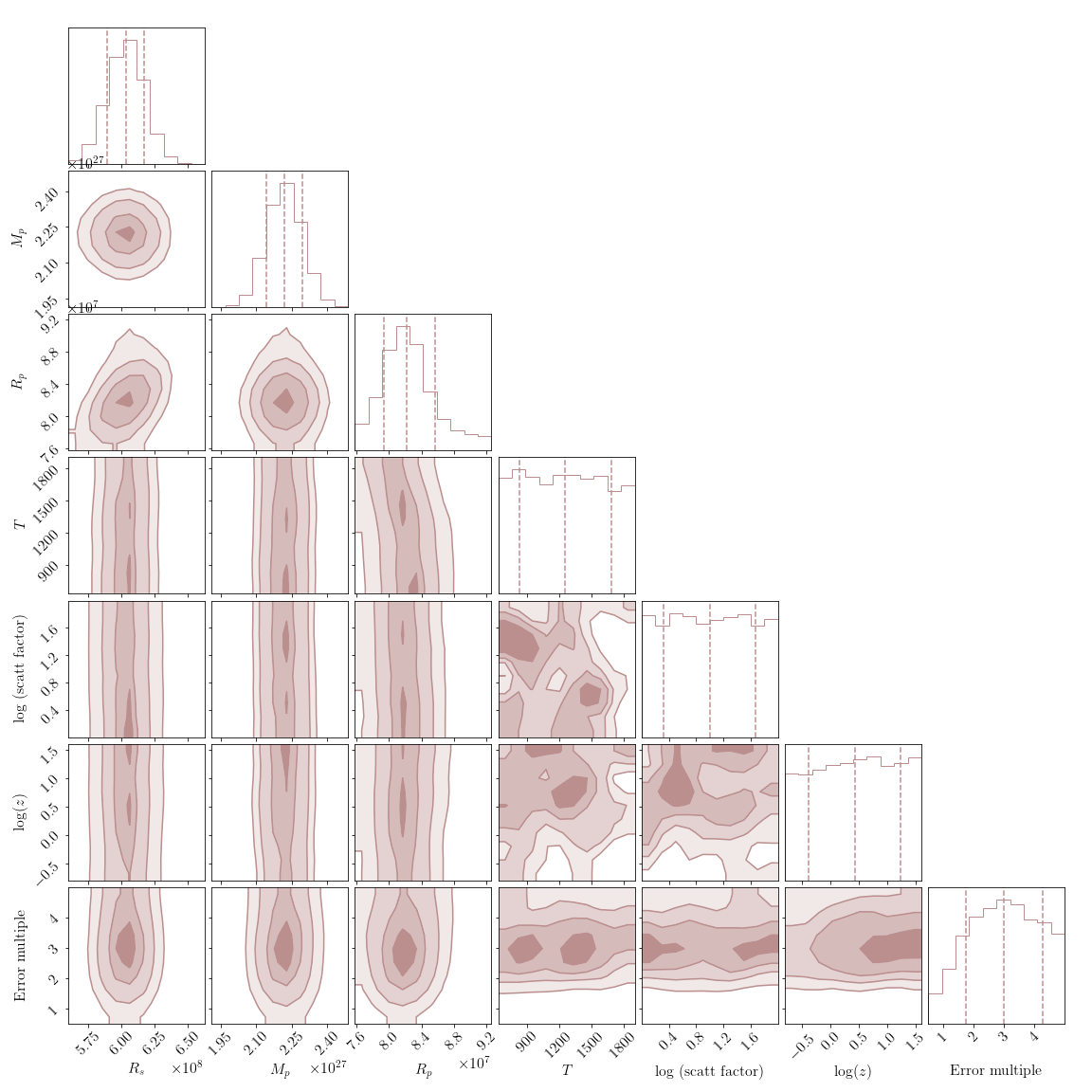}
\caption{Posterior probability distributions from \texttt{PLATON} retrieval for HAT-P-37b of wavelength range 501-779 nm (without $B$-filter)  .}
\label{fig:atmos_corner_VRri}
\end{center}
\end{figure*}


\bibliography{HATP37b}{}

\begin{thebibliography}{}
\expandafter\ifx\csname natexlab\endcsname\relax\def\natexlab#1{#1}\fi
\providecommand{\url}[1]{\href{#1}{#1}}
\providecommand{\dodoi}[1]{doi:~\href{http://doi.org/#1}{\nolinkurl{#1}}}
\providecommand{\doeprint}[1]{\href{http://ascl.net/#1}{\nolinkurl{http://ascl.net/#1}}}
\providecommand{\doarXiv}[1]{\href{https://arxiv.org/abs/#1}{\nolinkurl{https://arxiv.org/abs/#1}}}

\bibitem[{{Agol} {et~al.}(2005){Agol}, {Steffen}, {Sari}, \&
  {Clarkson}}]{agol2005}
{Agol}, E., {Steffen}, J., {Sari}, R., \& {Clarkson}, W. 2005, \mnras, 359,
  567, \dodoi{10.1111/j.1365-2966.2005.08922.x}

\bibitem[{{Awiphan} {et~al.}(2016){Awiphan}, {Kerins}, {Pichadee},
  {Komonjinda}, {Dhillon}, {Rujopakarn}, {Poshyachinda}, {Marsh}, {Reichart},
  {Ivarsen}, \& {Haislip}}]{awiphan2016}
{Awiphan}, S., {Kerins}, E., {Pichadee}, S., {et~al.} 2016, \mnras, 463, 2574,
  \dodoi{10.1093/mnras/stw2148}

\bibitem[{{Bakos} {et~al.}(2004){Bakos}, {Noyes}, {Kov{\'a}cs}, {Stanek},
  {Sasselov}, \& {Domsa}}]{bakos2004}
{Bakos}, G., {Noyes}, R.~W., {Kov{\'a}cs}, G., {et~al.} 2004, \pasp, 116, 266,
  \dodoi{10.1086/382735}

\bibitem[{{Bakos} {et~al.}(2012){Bakos}, {Hartman}, {Torres}, {B{\'e}ky},
  {Latham}, {Buchhave}, {Csubry}, {Kov{\'a}cs}, {Bieryla}, {Quinn},
  {Szklen{\'a}r}, {Esquerdo}, {Shporer}, {Noyes}, {Fischer}, {Johnson},
  {Howard}, {Marcy}, {Sato}, {Penev}, {Everett}, {Sasselov},
  {F{\H{u}}r{\'e}sz}, {Stefanik}, {L{\'a}z{\'a}r}, {Papp}, \&
  {S{\'a}ri}}]{bakos2012}
{Bakos}, G.~{\'A}., {Hartman}, J.~D., {Torres}, G., {et~al.} 2012, \aj, 144,
  19, \dodoi{10.1088/0004-6256/144/1/19}

\bibitem[{{Bakos} {et~al.}(2021){Bakos}, {Hartman}, {Bhatti}, {Csubry},
  {Penev}, {Bieryla}, {Latham}, {Quinn}, {Buchhave}, {Kov{\'a}cs}, {Torres},
  {Noyes}, {Falco}, {B{\'e}ky}, {Szklen{\'a}r}, {Esquerdo}, {Howard},
  {Isaacson}, {Marcy}, {Sato}, {Boisse}, {Santerne}, {H{\'e}brard}, {Rabus},
  {Harbeck}, {McCully}, {Everett}, {Horch}, {Hirsch}, {Howell}, {Huang},
  {L{\'a}z{\'a}r}, {Papp}, \& {S{\'a}ri}}]{bakos2021}
{Bakos}, G.~{\'A}., {Hartman}, J.~D., {Bhatti}, W., {et~al.} 2021, \aj, 162, 7,
  \dodoi{10.3847/1538-3881/abf637}

\bibitem[{{Bertin} \& {Arnouts}(1996)}]{berlin1996}
{Bertin}, E., \& {Arnouts}, S. 1996, \aaps, 117, 393,
  \dodoi{10.1051/aas:1996164}

\bibitem[{{Bonomo} {et~al.}(2017){Bonomo}, {Desidera}, {Benatti}, {Borsa},
  {Crespi}, {Damasso}, {Lanza}, {Sozzetti}, {Lodato}, {Marzari}, {Boccato},
  {Claudi}, {Cosentino}, {Covino}, {Gratton}, {Maggio}, {Micela}, {Molinari},
  {Pagano}, {Piotto}, {Poretti}, {Smareglia}, {Affer}, {Biazzo}, {Bignamini},
  {Esposito}, {Giacobbe}, {H{\'e}brard}, {Malavolta}, {Maldonado}, {Mancini},
  {Martinez Fiorenzano}, {Masiero}, {Nascimbeni}, {Pedani}, {Rainer}, \&
  {Scandariato}}]{bonomo2017}
{Bonomo}, A.~S., {Desidera}, S., {Benatti}, S., {et~al.} 2017, \aap, 602, A107,
  \dodoi{10.1051/0004-6361/201629882}

\bibitem[{{Borucki} {et~al.}(2010){Borucki}, {Koch}, {Basri}, {Batalha},
  {Brown}, {Caldwell}, {Caldwell}, {Christensen-Dalsgaard}, {Cochran},
  {DeVore}, {Dunham}, {Dupree}, {Gautier}, {Geary}, {Gilliland}, {Gould},
  {Howell}, {Jenkins}, {Kondo}, {Latham}, {Marcy}, {Meibom}, {Kjeldsen},
  {Lissauer}, {Monet}, {Morrison}, {Sasselov}, {Tarter}, {Boss}, {Brownlee},
  {Owen}, {Buzasi}, {Charbonneau}, {Doyle}, {Fortney}, {Ford}, {Holman},
  {Seager}, {Steffen}, {Welsh}, {Rowe}, {Anderson}, {Buchhave}, {Ciardi},
  {Walkowicz}, {Sherry}, {Horch}, {Isaacson}, {Everett}, {Fischer}, {Torres},
  {Johnson}, {Endl}, {MacQueen}, {Bryson}, {Dotson}, {Haas}, {Kolodziejczak},
  {Van Cleve}, {Chandrasekaran}, {Twicken}, {Quintana}, {Clarke}, {Allen},
  {Li}, {Wu}, {Tenenbaum}, {Verner}, {Bruhweiler}, {Barnes}, \&
  {Prsa}}]{borucki2010}
{Borucki}, W.~J., {Koch}, D., {Basri}, G., {et~al.} 2010, Science, 327, 977,
  \dodoi{10.1126/science.1185402}

\bibitem[{{Charbonneau} {et~al.}(2002){Charbonneau}, {Brown}, {Noyes}, \&
  {Gilliland}}]{charbon2002}
{Charbonneau}, D., {Brown}, T.~M., {Noyes}, R.~W., \& {Gilliland}, R.~L. 2002,
  \apj, 568, 377, \dodoi{10.1086/338770}

\bibitem[{{Czesla} {et~al.}(2019){Czesla}, {Schr{\"o}ter}, {Schneider},
  {Huber}, {Pfeifer}, {Andreasen}, \& {Zechmeister}}]{pya2019}
{Czesla}, S., {Schr{\"o}ter}, S., {Schneider}, C.~P., {et~al.} 2019, {PyA:
  Python astronomy-related packages}.
\newblock \doeprint{1906.010}

\bibitem[{{Deck} \& {Agol}(2016)}]{deck2016}
{Deck}, K.~M., \& {Agol}, E. 2016, \apj, 821, 96,
  \dodoi{10.3847/0004-637X/821/2/96}

\bibitem[{{Fabrycky} {et~al.}(2012){Fabrycky}, {Ford}, {Steffen}, {Rowe},
  {Carter}, {Moorhead}, {Batalha}, {Borucki}, {Bryson}, {Buchhave},
  {Christiansen}, {Ciardi}, {Cochran}, {Endl}, {Fanelli}, {Fischer}, {Fressin},
  {Geary}, {Haas}, {Hall}, {Holman}, {Jenkins}, {Koch}, {Latham}, {Li},
  {Lissauer}, {Lucas}, {Marcy}, {Mazeh}, {McCauliff}, {Quinn}, {Ragozzine},
  {Sasselov}, \& {Shporer}}]{fab2012}
{Fabrycky}, D.~C., {Ford}, E.~B., {Steffen}, J.~H., {et~al.} 2012, \apj, 750,
  114, \dodoi{10.1088/0004-637X/750/2/114}

\bibitem[{{Foreman-Mackey} {et~al.}(2013){Foreman-Mackey}, {Hogg}, {Lang}, \&
  {Goodman}}]{foreman2013}
{Foreman-Mackey}, D., {Hogg}, D.~W., {Lang}, D., \& {Goodman}, J. 2013, \pasp,
  125, 306, \dodoi{10.1086/670067}

\bibitem[{Fulton {et~al.}(2011)Fulton, Shporer, Winn, Holman, P{\'{a}}l, \&
  Gazak}]{Fulton2011}
Fulton, B.~J., Shporer, A., Winn, J.~N., {et~al.} 2011, The Astronomical
  Journal, 142, 84, \dodoi{10.1088/0004-6256/142/3/84}

\bibitem[{{Gim{\'e}nez} \& {Bastero}(1995)}]{gim1995}
{Gim{\'e}nez}, A., \& {Bastero}, M. 1995, \apss, 226, 99,
  \dodoi{10.1007/BF00626903}

\bibitem[{{Hayes} {et~al.}(2021){Hayes}, {Kerins}, {Morgan}, {Humpage},
  {Awiphan}, {Charles-Mindoza}, {McDonald}, {Inyanya}, {Padjaroen}, {Munsaket},
  {Chuanraksasat}, {Komonjinda}, {Kittara}, {Dhillon}, {Marsh}, {Reichart}, \&
  {Poshyachinda}}]{hayes2021}
{Hayes}, J.~J.~C., {Kerins}, E., {Morgan}, J.~S., {et~al.} 2021, arXiv
  e-prints, arXiv:2103.12139.
\newblock \doarXiv{2103.12139}

\bibitem[{{Holman} \& {Murray}(2005)}]{holman2005}
{Holman}, M.~J., \& {Murray}, N.~W. 2005, Science, 307, 1288,
  \dodoi{10.1126/science.1107822}

\bibitem[{{Husser} {et~al.}(2013){Husser}, {Wende-von Berg}, {Dreizler},
  {Homeier}, {Reiners}, {Barman}, \& {Hauschildt}}]{husser2013}
{Husser}, T.~O., {Wende-von Berg}, S., {Dreizler}, S., {et~al.} 2013, \aap,
  553, A6, \dodoi{10.1051/0004-6361/201219058}

\bibitem[{{Jiang} {et~al.}(2013){Jiang}, {Yeh}, {Thakur}, {Wu}, {Chien}, {Lin},
  {Chen}, {Hu}, {Sun}, \& {Ji}}]{jiang2013}
{Jiang}, I.-G., {Yeh}, L.-C., {Thakur}, P., {et~al.} 2013, \aj, 145, 68,
  \dodoi{10.1088/0004-6256/145/3/68}

\bibitem[{Kanodia \& Wright(2018)}]{Kano2018}
Kanodia, S., \& Wright, J. 2018, Research Notes of the {AAS}, 2, 4,
  \dodoi{10.3847/2515-5172/aaa4b7}

\bibitem[{{Kreidberg}(2015)}]{kreidberg2015}
{Kreidberg}, L. 2015, \pasp, 127, 1161, \dodoi{10.1086/683602}

\bibitem[{{Lang} {et~al.}(2010){Lang}, {Hogg}, {Mierle}, {Blanton}, \&
  {Roweis}}]{lang2010}
{Lang}, D., {Hogg}, D.~W., {Mierle}, K., {Blanton}, M., \& {Roweis}, S. 2010,
  \aj, 139, 1782, \dodoi{10.1088/0004-6256/139/5/1782}

\bibitem[{{Lithwick} {et~al.}(2012){Lithwick}, {Xie}, \& {Wu}}]{lith2012}
{Lithwick}, Y., {Xie}, J., \& {Wu}, Y. 2012, \apj, 761, 122,
  \dodoi{10.1088/0004-637X/761/2/122}

\bibitem[{{Maciejewski} {et~al.}(2010){Maciejewski}, {Dimitrov},
  {Neuh{\"a}user}, {Niedzielski}, {Raetz}, {Ginski}, {Adam}, {Marka},
  {Moualla}, \& {Mugrauer}}]{macie2010}
{Maciejewski}, G., {Dimitrov}, D., {Neuh{\"a}user}, R., {et~al.} 2010, \mnras,
  407, 2625, \dodoi{10.1111/j.1365-2966.2010.17099.x}

\bibitem[{{Maciejewski} {et~al.}(2016){Maciejewski}, {Dimitrov}, {Mancini},
  {Southworth}, {Ciceri}, {D'Ago}, {Bruni}, {Raetz}, {Nowak}, {Ohlert},
  {Puchalski}, {Saral}, {Derman}, {Petrucci}, {Jofre}, {Seeliger}, \&
  {Henning}}]{Maciejewski2016}
{Maciejewski}, G., {Dimitrov}, D., {Mancini}, L., {et~al.} 2016, \actaa, 66,
  55.
\newblock \doarXiv{1603.03268}

\bibitem[{{Maciejewski} {et~al.}(2018){Maciejewski}, {Fern{\'a}ndez},
  {Aceituno}, {Mart{\'\i}n-Ruiz}, {Ohlert}, {Dimitrov}, {Szyszka}, {von Essen},
  {Mugrauer}, {Bischoff}, {Michel}, {Mallonn}, {Stangret}, \&
  {Mo{\'z}dzierski}}]{macie2018}
{Maciejewski}, G., {Fern{\'a}ndez}, M., {Aceituno}, F., {et~al.} 2018, \actaa,
  68, 371, \dodoi{10.32023/0001-5237/68.4.4}

\bibitem[{{Mannaday} {et~al.}(2020){Mannaday}, {Thakur}, {Jiang}, {Sahu},
  {Joshi}, {Pandey}, {Joshi}, {Yadav}, {Su}, {Sariya}, {Yeh}, {Griv},
  {Mkrtichian}, {Shlyapnikov}, {Moskvin}, {Ignatov}, {Va{\v{n}}ko}, \&
  {P{\"u}sk{\"u}ll{\"u}}}]{mannaday2020}
{Mannaday}, V.~K., {Thakur}, P., {Jiang}, I.-G., {et~al.} 2020, \aj, 160, 47,
  \dodoi{10.3847/1538-3881/ab9818}

\bibitem[{{Parviainen} \& {Aigrain}(2015)}]{parviainen2015}
{Parviainen}, H., \& {Aigrain}, S. 2015, \mnras, 453, 3821,
  \dodoi{10.1093/mnras/stv1857}

\bibitem[{{Patra} {et~al.}(2017){Patra}, {Winn}, {Holman}, {Yu}, {Deming}, \&
  {Dai}}]{patra2017}
{Patra}, K.~C., {Winn}, J.~N., {Holman}, M.~J., {et~al.} 2017, \aj, 154, 4,
  \dodoi{10.3847/1538-3881/aa6d75}

\bibitem[{{Ricker} {et~al.}(2014){Ricker}, {Winn}, {Vanderspek}, {Latham},
  {Bakos}, {Bean}, {Berta-Thompson}, {Brown}, {Buchhave}, {Butler}, {Butler},
  {Chaplin}, {Charbonneau}, {Christensen-Dalsgaard}, {Clampin}, {Deming},
  {Doty}, {De Lee}, {Dressing}, {Dunham}, {Endl}, {Fressin}, {Ge}, {Henning},
  {Holman}, {Howard}, {Ida}, {Jenkins}, {Jernigan}, {Johnson}, {Kaltenegger},
  {Kawai}, {Kjeldsen}, {Laughlin}, {Levine}, {Lin}, {Lissauer}, {MacQueen},
  {Marcy}, {McCullough}, {Morton}, {Narita}, {Paegert}, {Palle}, {Pepe},
  {Pepper}, {Quirrenbach}, {Rinehart}, {Sasselov}, {Sato}, {Seager},
  {Sozzetti}, {Stassun}, {Sullivan}, {Szentgyorgyi}, {Torres}, {Udry}, \&
  {Villasenor}}]{ricker2014}
{Ricker}, G.~R., {Winn}, J.~N., {Vanderspek}, R., {et~al.} 2014, in Society of
  Photo-Optical Instrumentation Engineers (SPIE) Conference Series, Vol. 9143,
  Space Telescopes and Instrumentation 2014: Optical, Infrared, and Millimeter
  Wave, ed. J.~{Oschmann}, Jacobus~M., M.~{Clampin}, G.~G. {Fazio}, \& H.~A.
  {MacEwen}, 914320, \dodoi{10.1117/12.2063489}

\bibitem[{{Seager} \& {Sasselov}(2000)}]{seager2000}
{Seager}, S., \& {Sasselov}, D.~D. 2000, \apj, 537, 916, \dodoi{10.1086/309088}

\bibitem[{{Sing} {et~al.}(2016){Sing}, {Fortney}, {Nikolov}, {Wakeford},
  {Kataria}, {Evans}, {Aigrain}, {Ballester}, {Burrows}, {Deming},
  {D{\'e}sert}, {Gibson}, {Henry}, {Huitson}, {Knutson}, {Lecavelier Des
  Etangs}, {Pont}, {Showman}, {Vidal-Madjar}, {Williamson}, \&
  {Wilson}}]{sing2016}
{Sing}, D.~K., {Fortney}, J.~J., {Nikolov}, N., {et~al.} 2016, \nat, 529, 59,
  \dodoi{10.1038/nature16068}

\bibitem[{{Southworth} {et~al.}(2019){Southworth}, {Dominik}, {J{\o}rgensen},
  {Andersen}, {Bozza}, {Burgdorf}, {D'Ago}, {Dib}, {Figuera Jaimes}, {Fujii},
  {Gill}, {Haikala}, {Hinse}, {Hundertmark}, {Khalouei}, {Korhonen},
  {Longa-Pe{\~n}a}, {Mancini}, {Peixinho}, {Rabus}, {Rahvar}, {Sajadian},
  {Skottfelt}, {Snodgrass}, {Spyratos}, {Tregloan-Reed}, {Unda-Sanzana}, \&
  {von Essen}}]{southworth2019}
{Southworth}, J., {Dominik}, M., {J{\o}rgensen}, U.~G., {et~al.} 2019, \mnras,
  490, 4230, \dodoi{10.1093/mnras/stz2602}

\bibitem[{{Speagle}(2020)}]{speagle2020}
{Speagle}, J.~S. 2020, \mnras, 493, 3132, \dodoi{10.1093/mnras/staa278}

\bibitem[{{Stassun} {et~al.}(2019){Stassun}, {Oelkers}, {Paegert}, {Torres},
  {Pepper}, {De Lee}, {Collins}, {Latham}, {Muirhead}, {Chittidi},
  {Rojas-Ayala}, {Fleming}, {Rose}, {Tenenbaum}, {Ting}, {Kane}, {Barclay},
  {Bean}, {Brassuer}, {Charbonneau}, {Ge}, {Lissauer}, {Mann}, {McLean},
  {Mullally}, {Narita}, {Plavchan}, {Ricker}, {Sasselov}, {Seager}, {Sharma},
  {Shiao}, {Sozzetti}, {Stello}, {Vanderspek}, {Wallace}, \&
  {Winn}}]{stassun2019}
{Stassun}, K.~G., {Oelkers}, R.~J., {Paegert}, M., {et~al.} 2019, \aj, 158,
  138, \dodoi{10.3847/1538-3881/ab3467}

\bibitem[{{Turner} {et~al.}(2017){Turner}, {Leiter}, {Biddle}, {Pearson},
  {Hardegree-Ullman}, {Thompson}, {Teske}, {Cates}, {Cook}, {Berube},
  {Nieberding}, {Jones}, {Raphael}, {Wallace}, {Watson}, \&
  {Johnson}}]{turner2017}
{Turner}, J.~D., {Leiter}, R.~M., {Biddle}, L.~I., {et~al.} 2017, \mnras, 472,
  3871, \dodoi{10.1093/mnras/stx2221}

\bibitem[{{von Essen} {et~al.}(2019){von Essen}, {Wedemeyer}, {Sosa}, {Hjorth},
  {Parkash}, {Freudenthal}, {Mallonn}, {Micul{\'a}n}, {Zibecchi}, {Cellone}, \&
  {Torres}}]{von2019}
{von Essen}, C., {Wedemeyer}, S., {Sosa}, M.~S., {et~al.} 2019, \aap, 628,
  A116, \dodoi{10.1051/0004-6361/201731966}

\bibitem[{{Wang} {et~al.}(2021){Wang}, {Wang}, {Wang}, {Wu}, {Rice}, {Zhou},
  {Hinse}, {Liu}, {Ma}, {Peng}, {Zhang}, {Yu}, {Zhou}, \&
  {Laughlin}}]{wang2021}
{Wang}, X.-Y., {Wang}, Y.-H., {Wang}, S., {et~al.} 2021, \apjs, 255, 15,
  \dodoi{10.3847/1538-4365/ac0835}

\bibitem[{{Yang} {et~al.}(2021){Yang}, {Wang}, {Li}, \& {Liu}}]{yang2021}
{Yang}, J.-M., {Wang}, X.-Y., {Li}, K., \& {Liu}, Y. 2021, \pasj,
  \dodoi{10.1093/pasj/psab059}

\bibitem[{{Zechmeister} \& {K{\"u}rster}(2009)}]{zech2009}
{Zechmeister}, M., \& {K{\"u}rster}, M. 2009, \aap, 496, 577,
  \dodoi{10.1051/0004-6361:200811296}

\bibitem[{{Zhang} {et~al.}(2019){Zhang}, {Chachan}, {Kempton}, \&
  {Knutson}}]{zhang2019}
{Zhang}, M., {Chachan}, Y., {Kempton}, E. M.~R., \& {Knutson}, H.~A. 2019,
  \pasp, 131, 034501, \dodoi{10.1088/1538-3873/aaf5ad}

\end{thebibliography}
\bibliographystyle{aasjournal}



\end{document}